%% file: arxiv_absorbing_centrality.tex
\documentclass[conference]{IEEEtran}

\input{macros}

\title{Absorbing random-walk centrality: \\ 
Theory and algorithms}

\IEEEoverridecommandlockouts 

\author{
\IEEEauthorblockN{Charalampos Mavroforakis\\}
\IEEEauthorblockA{Dept.\ of Computer Science\\
Boston University\\
Boston, U.S.A.\\
cmav@cs.bu.edu\\}
\and
\IEEEauthorblockN{Michael Mathioudakis and Aristides Gionis}
\IEEEauthorblockA{Helsinki Institute for Information Technology HIIT\\
Dept.\ of Computer Science, Aalto University\\
Helsinki, Finland \\
firstname.lastname@aalto.fi\\}
}

\begin{document}
\maketitle

\begin{abstract}
We study a new notion of graph centrality
based on absorbing random walks.  
Given a graph $G=(V,E)$ and a set of {\em query nodes} $Q\subseteq V$,
we aim to identify the $k$ most central nodes in $G$ with respect to~$Q$.
Specifically, we consider central nodes to be {\em absorbing} 
for random walks that start at the query nodes~$Q$. 
The goal is to find the set of $k$ central nodes
that minimizes the expected length of a random walk until
absorption. 
The proposed measure, which we call 
$k$ {\em absorbing random-walk centrality},  
favors {\em diverse} sets, 
as it is beneficial to place the $k$ absorbing nodes in different parts of the
graph so as to ``intercept'' random walks that start from different query nodes.

\smallskip
Although similar problem definitions have been considered in the
literature, e.g., in information-ret\-rieval settings
where the goal is to diversify web-search results, 
in this paper we study the problem formally and prove some of its properties. 
We show that the problem is \NP-hard, while the objective function is
monotone and supermodular, implying that a greedy algorithm
provides solutions with an approximation guarantee.
On the other hand, the greedy algorithm involves expensive
matrix operations that make it prohibitive to employ on large datasets.
To confront this challenge, we develop more efficient
algorithms based on spectral clustering and on personalized PageRank. 
\end{abstract}


\section{Introduction}

A fundamental problem in graph mining is to
identify the most central nodes in a graph.
Numerous centrality measures have been proposed, 
including
degree centrality, 
closeness centrality~\cite{closenesscentrality}, 
betweenness centrality~\cite{betweennesscentrality}, 
random-walk centrality~\cite{randomwalkcentrality},
Katz centrality~\cite{katzcentrality}, and 
PageRank~\cite{PageRank}.

In the interest of robustness
many centrality measures use random walks:
while the shortest-path distance between two nodes can
change dramatically by inserting or deleting a single edge, 
distances based on random walks account for multiple paths and offer a
more global view of the connectivity between two nodes. 
In this spirit, the random-walk centrality of {\it one} node with
respect to {\it all nodes} of the graph is defined as the
expected time needed to come across this node in a random walk that
starts in any other node of the graph~\cite{randomwalkcentrality}.

In this paper, we consider a measure that generalizes random-walk
centrality for a {\it set} of nodes $C$
with respect to a set of \emph{query nodes} $Q$.
Our centrality measure is defined as the expected length
of a random walk that starts from any node in $Q$ until
it reaches any node in $C$ --- at which point the random walk
is {\it ``absorbed''} by $C$. Moreover, to allow for adjustable
importance of query nodes in the centrality measure,
we consider random walks {\it with restarts}, that occur with a fixed
probability $\alpha$ at each step of the random walk.
The resulting computational problem
is to find a set of $k$ nodes $C$ that optimizes this measure with respect to nodes $Q$,
which are provided as input. 
We call this measure  $k$ {\em absorbing random-walk centrality}
and the corresponding optimization problem {\krwc}.

To motivate the \krwc\ problem, let us consider the scenario of
searching the Web graph and summarizing the search results.
In this scenario, nodes of the graph correspond to webpages,
edges between nodes correspond to links between pages, and the set of
query nodes $Q$ consists of all nodes that match a user query, 
i.e.,  all webpages that satisfy a keyword search. 
Assuming that the size of $Q$ is large, the goal is to find the $k$
most central nodes with respect to $Q$, 
and present those to the user.

It is clear that ordering the nodes of the graph by their individual
random-walk centrality scores and taking the top-$k$ set does not solve the
\krwc\ problem, 
as these nodes may all be located in the
same ``neighborhood'' of the graph, and thus, 
may not provide a good absorbing set for the query.
On the other hand, 
as the goal is to minimize the expected absorption time for walks starting at $Q$, 
the optimal solution to the \krwc\ problem will be a set of $k$, both
{\em centrally-placed} and {\em diverse}, nodes.

This observation has motivated researchers in the informa\-tion-retrieval
field to consider random walks with absorbing states in order
to diversify web-search results~\cite{Zhu:2007tj}. 
However, despite the fact that similar problem definitions and algorithms have
been considered earlier, the \krwc\ problem has not been formally
studied and there has not been a theoretical analysis of its properties. 

Our key results in this paper are the following:
we show that the \krwc\ problem is \NP-hard, and 
we show that the $k$ absorbing random-walk centrality measure is 
{\em monotone} and {\em supermodular}.  
The latter property allows us to quantify the approximation guarantee
obtained by a natural greedy algorithm, 
which has also been considered by previous work~\cite{Zhu:2007tj}. 
Furthermore, a na\"{i}ve implementation of the greedy algorithm requires 
many expensive matrix inversions, which make the algorithm particularly slow. 
Part of our contribution is to show how to make use of the 
Sherman-Morrison inversion formula to implement the greedy algorithm 
with only one matrix inversion and more efficient
matrix$\,\times\,$vector multiplications.

Moreover, we explore the performance of faster, heuristic algorithms,
aiming to identify methods that are faster than the greedy approach
without significant loss in the quality of results. The heuristic 
algorithms we consider include the personalized PageRank
algorithm~\cite{PageRank,langville2005survey}
as well as algorithms based on spectral clustering~\cite{von2007tutorial}. We find that, in
practice, the personalized PageRank algorithm offers a very good trade-off
between speed and quality.

The rest of the paper is organized as follows. 
In Section~\ref{sec:related},
we overview previous work and discuss how it compares to this paper.
We define our problem in Section~\ref{section:problem-definition}
and provide basic background results on absorbing random walks in
Section~\ref{section:absorbing}. 
Our main technical contributions are given in Sections~\ref{section:absorbing} and~\ref{sec:properties}, 
where we characterize the complexity of the problem, and provide the details of the
greedy algorithm and the heuristics we explore. 
We evaluate the performance of algorithms in Section~\ref{sec:experiments},
over a range of real-world graphs, 
and Section~\ref{sec:conclusions} is a short conclusion.
Proofs for some of the theorems shown in the paper are
provided in the Appendix.

\section{Related work}
\label{sec:related}

Many works in the literature explore ways to quantify
the notion of node centrality on graphs~\cite{boldi2014axioms}.
Some of the most commonly-used measures include the following:
($i$) {\em degree centrality}, where the centrality of a node is
simply quantified by its degree; 
($ii$) {\em closeness centrality} \cite{leavitt1951some,closenesscentrality}, defined
as the average distance of a node from all other nodes on the
graph;
($iii$) {\em betweenness centrality}~\cite{betweennesscentrality},
defined as the number of shortest paths between pairs of
nodes in the graph that pass through a given node; 
($iv$) {\em eigenvector centrality}, defined as the stationary probability
that a Markov chain on the graph visits a given node, with
Katz centrality~\cite{katzcentrality} and PageRank~\cite{PageRank}
being two well-studied variants; and 
($v$) {\em random-walk centrality}~\cite{randomwalkcentrality}, defined as
the expected first passage time of a random walk from a given node,
when it starts from a random node of the graph.
The measure we study in this paper
generalizes the notion of {\em random-walk centrality} to a set
of absorbing nodes.


Absorbing random walks have been used in previous work
to select a {\it diverse} set of nodes from a graph. 
For example,
an algorithm proposed by Zhu et~al.~\cite{Zhu:2007tj} selects nodes in the following manner: 
($i$) the first node is selected based on its PageRank value and is set as absorbing; 
($ii$) the next node to be selected is the node that maximizes the expected first-passage time from the already selected absorbing nodes.
Our problem definition differs considerably from the one considered in that work, 
as in our work the expected first-passage times are always computed from the set of
query nodes that are provided in the input, and not from the nodes that participate in the solution so far. 
In this respect, the greedy method proposed by Zhu et~al.\ is not associated with a crisp problem definition.

Another conceptually related line of work aims to
select a diverse subset of query results, mainly
within the context of document retrieval~\cite{Agrawal:2009bc,Angel:2011be,Vieira:2011bx}. 
The goal, there, is to select $k$ query results to
optimize a function that quantifies the trade-off between relevance
and diversity.

Our work is also remotely related to the problem studied by Leskovec et al.\ on
{\it cost-effective outbreak detection} \cite{leskovec2007cost}. 
One of the problems discussed there is to select
nodes in the network so that the detection time for a set of cascades is minimized. 
However, their work differs from ours on the fact that they consider as input a set of {\it cascades}, 
each one of finite size, 
while in our case the input consists of a set of query {\it nodes} 
and we consider a probabilistic model that generates random walk paths, 
of possibly infinite size.

\section{Problem definition}
\label{section:problem-definition}

We are given a graph $G=(V,E)$ over a set of nodes $V$ and set of
undirected edges $E$. The number of nodes $|V|$ is denoted by $n$ and the
number of edges $|E|$ by $m$. The input also includes a subset of
nodes $Q\subseteq V$, to which we refer as the {\em query nodes}.
As a special case, the set of query nodes $Q$ may be equal to the
whole set of nodes, i.e., $Q=V$.

Our goal is to find a set $C$ of $k$
nodes that are {\em central} with respect to the query nodes $Q$. 
For some applications it makes sense to restrict the central nodes to 
be only among the query nodes, while in other cases, 
the central nodes may include any node in $V$.
To model those different scenarios, we consider a set of candidate nodes $D$, 
and require that the $k$ central nodes should belong in this candidate set, 
i.e., $C\subseteq D$. 
Some of the cases include $D=Q$, $D=V$, or $D=V\setminus Q$, but
it could also be that $D$ is defined in some other way
that does not involve $Q$. 
In general, we assume that $D$ is given as input.

The {\em centrality} of a set of nodes $C$ with respect to query nodes $Q$
is based on the notion of absorbing random-walks and their expected
length.
More specifically, let us consider a random walk on the nodes $V$ of
the graph, that proceeds at discrete steps: the walk starts from a
node $q\in Q$ and, at each step 
moves to a different node, following edges in $G$, 
until it arrives at some node in $C$.
The {\em starting} node $q$ of the walk is chosen according to a probability distribution $\startprob$.
When the walk arrives at a node $c\in C$ for the first time,
it terminates, and we say that the random walk is {\em absorbed} by that node~$c$.
In the interest of generality, and to allow for adjustable
importance of query nodes in the centrality measure,
we also allow the random walk to restart. Restarts occur with a
probability $\alpha$ at each step of the random walk, where $\alpha$ is 
a parameter that is specified as input to the problem.
When restarting, the walk proceeds to a query node selected randomly
according to $\startprob$.
Intuitively, larger values of $\alpha$ favor nodes that are closer to nodes $Q$.

We are interested in the expected length (i.e., number of steps)
of the walk that starts from a query node $q\in Q$ until it gets
absorbed by some node in~$C$, and we denote this expected length
by $\abscentrality^q_{_Q}(C)$. 
We then define the {\em absorbing random-walk centrality} of a set of
nodes $C$ with respect to query nodes~$Q$,~by 
\[
\abscentrality_Q(C) = \sum_{q\in Q} \startprob(q) \, \abscentrality^q_{_Q}(C).
\]

The problem we consider in this paper is the following. 
\begin{problem}
{\em (}\krwc{\em )}
\label{problem:k-ac}
We are given a graph $G=(V,E)$, 
a set of query nodes $Q \subseteq V$, 
a set of candidate nodes $D\subseteq V$,
a starting probability distribution $\startprob$ over $V$ such that $\startprob(v)=0$ if $v \in V \setminus Q$,
a restart probability~$\alpha$, 
and an integer $k$.
We ask to
find a set of $k$ nodes $C\subseteq D$ 
that {\em minimizes} $\abscentrality_Q(C)$, i.e., 
the expected length of a random walk that starts from $Q$ and proceeds
until it gets absorbed in some node in $C$.
\end{problem}
In cases where we have no reason to distinguish among the query nodes,
we consider the uniform starting probability distribution 
$\startprob(q) = 1/|Q|$.
In fact, for simplicity of exposition, hereinafter we focus on the
case of uniform distribution. However, we note that all our definitions
and techniques generalize naturally, not only to general starting 
probability distributions $\startprob(q)$, but also to 
{\em directed} and {\em weighted} graphs.

\section{Absorbing random walks}
\label{section:absorbing}

In this section we review some relevant background on absorbing random
walks. Specifically, we discuss how to calculate the objective function 
$\abscentrality_Q(C)$ for Problem~\ref{problem:k-ac}.

Let  $\mathbf{P}$ be the transition matrix for a random walk,
with $\mathbf{P}(i,j)$ expressing the probability that the random
walk will move to node~$j$ given that it is currently at node $i$.
Since random walks can only move to absorbing nodes $C$, but not away
from them, we set 
$\mathbf{P}(c,c)=1$ and 
$\mathbf{P}(c,j)=0$, if $j\neq c$, 
for all absorbing nodes $c\in C$.
The set $T = V \setminus C$ of non-absorbing nodes is called {\em transient}.
If $N(i)$ are the neighbors
of a node $i \in T$ and $d_i = |N(i)|$ its degree, the transition
probabilities from node $i$ to other nodes are
\begin{equation}
\mathbf{P}(i,j) = \left\{
\begin{array}{ll}
\alpha \, \startprob(j) & \text{ if } j \in Q \setminus N(i), \\
(1-\alpha) / d_i+ \alpha \, \startprob(j)  & \text{ if } j \in N(i).
\end{array}
\right.
\end{equation}
Here, $\startprob$ represents the starting probability vector.
For example, for the uniform distribution over query nodes we have
$\startprob(i) = 1/|Q|$ if $i\in Q$ and $0$ otherwise. 
The transition matrix of
the random walk can be
written as follows
\begin{equation}
\mathbf{P} = \left(
\begin{array}{cc}
\mathbf{P}_{TT} & \mathbf{P}_{TC} \\
\mathbf{0} & \mathbf{I} 
\end{array}
\right).
\end{equation}
In the equation above, $\mathbf{I}$ is an $(n - |T|)\times(n - |T|)$
identity matrix and $\mathbf{0}$ a matrix with all its entries equal
to $0$; 
$\mathbf{P}_{TT}$ is the $|T| \times |T|$ sub-matrix of $\mathbf{P}$
that contains the transition probabilities between transient nodes; and 
$\mathbf{P}_{TC}$ is the $|T|\times|C|$ sub-matrix of $\mathbf{P}$
that contains the transition probabilities from transient to absorbing nodes.

The probability of the walk being on node $j$ at exactly
$\ell$ steps having started at node~$i$, 
is given by the $(i,j)$-entry of the matrix $\mathbf{P}_{TT}^\ell$. 
Therefore, the expected total number of times that the random walk visits
node $j$ having started from node~$i$ is given by the $(i,j)$-entry of 
the $|T|\times |T|$ matrix
\begin{equation}
\mathbf{F} = \sum_{\ell=0}^{\infty} \mathbf{P}_{TT}^\ell 
= \left( \mathbf{I} - \mathbf{P}_{TT}\right)^{-1}, 
\end{equation}
which is known as the {\em fundamental matrix} of the
absorbing random walk. 
Allowing the possibility to start the random walk at an absorbing node 
(and being absorbed immediately), we see that the expected length of a
random walk that starts from node $i$ and gets absorbed by the set $C$
is given by the $i$-th element of the following $n\times 1$ vector
\begin{equation}
\mathbf{L} = \mathbf{L}_C = \left(
\begin{array}{c}
\mathbf{F} \\ 
\mathbf{0}
\end{array}
\right) \mathbf{1}, 
\end{equation}
where $\mathbf{1}$ is an $T\times 1$ vector of all 1s.
We write $\mathbf{L}=\mathbf{L}_C$ to emphasize the dependence on the
set of absorbing nodes $C$.

The expected number of steps when starting from a node in $Q$ and
until being absorbed by some node in $C$ is then obtained by summing
over all query nodes, i.e., 
\begin{equation}
\label{eq:inversion}
\abscentrality_Q(C) =  \startprob^T\, \mathbf{L}_C.
\end{equation}

\subsection{Efficient computation of absorbing centrality}
\label{section:fasteval}

Equation~(\ref{eq:inversion}) pinpoints the difficulty of the problem we consider: 
even computing the objective function $\abscentrality_Q(C)$ for a
candidate solution $C$ requires an expensive matrix inversion;
$\mathbf{F} = \left( \mathbf{I} - \mathbf{P}_{TT}\right)^{-1}$.
Furthermore, 
searching for the optimal set $C$ 
involves an exponential number of candidate sets, 
while evaluating each one of them requires a matrix inversion.

In practice, we find that we can compute $\abscentrality_Q(C)$ much faster approximately, as shown in Algorithm~\ref{algo:appox_ac}.
The algorithm follows from the infinite-sum expansion of Equation~(\ref{eq:inversion}).
\begin{eqnarray*}
\abscentrality_Q(C) =  \startprob^T\, \mathbf{L}_C = \startprob^T\, 
\left(
\begin{array}{c}
\mathbf{F} \\ 
\mathbf{0}
\end{array}
\right) \mathbf{1} 
= \startprob^T\, 
\left(
\begin{array}{c}
\sum_{\ell=0}^{\infty}\mathbf{P}_{TT}^\ell \\ 
\mathbf{0}
\end{array}
\right) \mathbf{1} \\ 
= \startprob^T\,
\sum_{\ell=0}^{\infty} 
\left(
\begin{array}{c}
\mathbf{P}_{TT}^\ell \\ 
\mathbf{0}
\end{array}
\right) \mathbf{1}
= \left(\sum_{\ell=0}^{\infty}  \startprob^T\,
\left(
\begin{array}{c}
\mathbf{P}_{TT}^\ell \\ 
\mathbf{0}
\end{array}
\right)\right) \mathbf{1} \\ 
= \left(\sum_{\ell=0}^{\infty}  \mathbf{x}_\ell \right) \mathbf{1}
= \sum_{\ell=0}^{\infty}  \mathbf{x}_\ell \mathbf{1},
\end{eqnarray*}
with
\begin{equation}
\mathbf{x}_0 = \startprob^{^T} \;\text{ and }\; \mathbf{x}_{\ell+1} = \mathbf{x}_\ell \left(
\begin{array}{c}
\mathbf{P}_{TT} \\ 
\mathbf{0}
\end{array}
\right).
\end{equation}
Note that computing each vector $\mathbf{x}_\ell$ requires time~$\mathcal{O}(n^2)$.
Algorithm~\ref{algo:appox_ac} terminates 
when the increase of the sum 
due to the latest term 
falls below a pre-defined threshold $\epsilon$.

\begin{algorithm}[t]
\caption{\approxac}
\label{algo:appox_ac}
\begin{algorithmic}
\STATE {\bf Input}: Transition matrix $\mathbf{P}_{TT}$, threshold $\epsilon$, 
\\ starting probabilities $\startprob$
\STATE {\bf Output}: Absorbing centrality $\abscentrality_Q$
\STATE $\mathbf{x_0} \leftarrow \startprob^{^T}$ 
\STATE $\delta \leftarrow \mathbf{x_0}\cdot\mathbf{1}$ 
\STATE $\abscentrality \leftarrow  \delta$ 
\STATE $\ell \leftarrow 0$ 
\WHILE {$\delta < \epsilon$}
\STATE $\mathbf{x_{\ell + 1}} \leftarrow \mathbf{x_\ell} \left(
\begin{array}{c}
\mathbf{P}_{TT} \\ 
\mathbf{0}
\end{array}
\right)$ 
\STATE $\delta \leftarrow \mathbf{x_{\ell + 1}}\cdot \mathbf{1}$ 
\STATE $\abscentrality \leftarrow \abscentrality + \delta$ 
\STATE $\ell \leftarrow \ell + 1$
\ENDWHILE
\STATE \textbf{return} \abscentrality
\end{algorithmic}
\end{algorithm}

\section{Problem characterization}
\label{sec:properties}

We now study the \krwc\ problem in more detail.
In particular, we show that the function $\abscentrality_Q$
is monotone and supermodular, a property that is used later to provide
an approximation guarantee for the greedy algorithm.
We also show that \krwc\ is  \NP-hard. 

Recall that a function $f:2^V\rightarrow\mathbb{R}$ over subsets of
a ground set $V$ is {\em submodular} if it has the 
{\em diminishing returns} property
\begin{equation}
f(Y \cup \{u\}) - f(Y) \le f(X \cup \{u\}) - f(X), 
\end{equation}
for all $X\subseteq Y\subseteq V$ and $u\not\in Y$.
The function $f$ is {\em supermodular} if $-f$ is submodular. 
Submodularity (and supermodularity) is a very useful property for
designing algorithms. 
For instance, minimizing a submodular function is a polynomial-time solvable problem, 
while the maximization problem is typically amenable to approximation algorithms, 
the exact guarantee of which depends on other properties of the function and requirements of the problem,
e.g., monotonicity, matroid constraints, etc.

Even though the objective function $\abscentrality_Q(C)$ is
given in closed-form by Equation~(\ref{eq:inversion}),
to prove its properties we find it more convenient to work with its
descriptive definition, namely, 
$\abscentrality_Q(C)$ being the expected length 
for a random walk starting at nodes of $Q$ 
before being absorbed at nodes of~$C$.

For the rest of this section we consider that the set of query nodes
$Q$ is fixed, and for simplicity we write
$\abscentrality=\abscentrality_Q$.

\begin{proposition}[Monotonicity]
\label{proposition:monotonicity}
For all $X\subseteq Y\subseteq V$ it is
$\abscentrality(Y)\le\abscentrality(X)$.
\end{proposition}

The proposition states that absorption time decreases with more absorbing
nodes. The proof is given in the Appendix.

Next we show that the absorbing random-walk centrality measure
$\abscentrality(\cdot)$ is supermodular.

\begin{proposition}[Supermodularity]
\label{proposition:supermodularity}
For all sets $X\subseteq Y\subseteq V$ and $u \not\in Y$ it is
\begin{equation}
\label{eq:supermodularity}
\abscentrality(X) - \abscentrality(X \cup \{u\}) \ge  
\abscentrality(Y) - \abscentrality(Y \cup \{u\}) .
\end{equation}
\end{proposition}

\begin{IEEEproof}
Given an instantiation of a random walk, we define the following propositions
for any pair of nodes $i, j\in V$, non-negative integer $\ell$,
and set of nodes $Z$:
\begin{description}
\item[$A_{i,j}^\ell(Z)$:] The random walk started at node $i$ and
  visited node~$j$ after exactly $\ell$ steps, without visiting  any
  node in set~$Z$. 
\item[$B_{i,j}^\ell(Z, u)$:] The random walk started at node $i$ and 
  visited node~$j$ after exactly $\ell$ steps, having previously visited
  node~$u$ but without visiting any node in the set~$Z$.
\end{description}
It is easy to see that the set of random walks for which
$A_{i,j}^\ell(Z)$ is \true\ can be partitioned into
those that visited $u$ within the first $\ell$ steps and those
that did not.
Therefore, the probability that proposition $A_{i,j}^\ell(Z)$
is \true\ for any instantiation of a random walk generated by our model is equal to
\begin{equation}
\label{eq:split_paths_probability}
\Pr \left[ A_{i,j}^\ell(Z) \right] = 
\Pr \left[A_{i,j}^\ell(Z \cup \{u\})\right] + \Pr \left[B_{i,j}^\ell(Z,u)\right].
\end{equation}
Now, let $\stepstoZ(Z)$ be the number of steps for a random walk to reach the nodes in $Z$. 
$\stepstoZ(Z)$ is a random variable and its expected value
over all random walks generated by our model is equal
to $\abscentrality(Z)$.
Note that the proposition $\stepstoZ(Z) \geq \ell + 1$
is \true\ for a given instantiation of a random walk only 
if there is a pair of nodes $q\in Q$ and $j\in V\setminus Z$,
for which the proposition $A^\ell_{q,j}(Z)$ is \true.
Therefore,
\begin{equation}
\Pr \left[ \stepstoZ(Z) \geq \ell + 1 \right] 
= \sum_{q \in Q} \sum_{j \in V \setminus Z} \Pr \left[ A_{q,j}^\ell(Z)\right].	
\end{equation}
From the above, it is easy to calculate $\abscentrality(Z)$ as
\begin{eqnarray}
\abscentrality(Z) & = & E[\stepstoZ(Z)] \nonumber  \nonumber \\ 
& = & \sum_{\ell = 0}^{\infty} \ell \, \Pr \left[ \stepstoZ(Z) = \ell \right]  \nonumber \\ 
& = & \sum_{\ell = 1}^{\infty} \Pr \left[ \stepstoZ(Z) \geq \ell \right]  \nonumber \\
& = & \sum_{\ell = 0}^{\infty} \Pr \left[ \stepstoZ(Z) \geq \ell + 1 \right]  \nonumber \\
& = & \sum_{\ell = 0}^{\infty} \sum_{q \in Q} \sum_{j \in V \setminus Z} \Pr \left[ A_{q,j}^\ell(Z)\right] .
\label{equation:expectation}
\end{eqnarray}
The final property we will need is the observation that, 
for $X \subseteq Y$, 
$B_{i,j}^\ell (Y,u)$ implies $B_{i,j}^\ell (X,u)$  and thus
\begin{equation}
\label{eq:Beta_event_probability}
\Pr \left[ B_{i,j}^\ell (X,u) \right] \ge \Pr \left[ B_{i,j}^\ell (Y,u) \right].
\end{equation}
By using Equation~(\ref{equation:expectation}), 
the Inequality~(\ref{eq:supermodularity}) can be rewritten as 
\begin{align}
\nonumber
\sum_{\ell=0}^{\infty} \sum_{q \in Q} & \sum_{j \in V \setminus X} \Pr
\left[ A_{q,j}^\ell(X)\right] - \\
\nonumber
& \sum_{\ell=0}^{\infty} \sum_{q \in Q} \sum_{j \in V \setminus \{X
  \cup \{u\}\}} \Pr \left[ A_{q,j}^\ell(X \cup \{u\})\right] \\  
\nonumber
\ge \sum_{\ell=0}^{\infty} \sum_{q \in Q} & \sum_{j \in V \setminus Y}
\Pr \left[ A_{q,j}^\ell(Y)\right] - \\
& \sum_{\ell=0}^{\infty} \sum_{q \in Q} \sum_{j \in V \setminus \{Y \cup \{u\}\}} \Pr \left[ A_{q,j}^\ell(Y \cup \{u\})\right].
\end{align}
We only need to show that the inequality holds for an arbitrary value of $\ell$ and $q \in Q$, 
that is
\begin{align}
\nonumber
&\sum_{j \in V \setminus X} \Pr \left[ A_{q,j}^\ell(X)\right] 
- \sum_{j \in V \setminus \{X \cup \{u\}\}} \Pr \left[ A_{q,j}^\ell(X \cup \{u\})\right]  \ge \\
& \sum_{j \in V \setminus Y} \Pr \left[ A_{q,j}^\ell(Y)\right] 
- \sum_{j \in V \setminus \{Y \cup \{u\}\}} \Pr \left[ A_{q,j}^\ell(Y \cup \{u\})\right].
\end{align}
Notice that $\Pr \left[ A_{i,u}^\ell(Y \cup \{u\}) \right] = 0$, 
so we can rewrite the above inequality as 
\begin{align}
\nonumber
&\sum_{j \in V \setminus X} \Pr \left[ A_{q,j}^\ell(X)\right] 
- \sum_{j \in V \setminus X } \Pr \left[ A_{q,j}^\ell(X \cup \{u\})\right] \ge  \\
&\sum_{j \in V \setminus Y} \Pr \left[ A_{q,j}^\ell(Y)\right] 
- \sum_{j \in V \setminus Y } \Pr \left[ A_{q,j}^\ell(Y \cup \{u\})\right].
\end{align}
To show the latter inequality we start from the left hand side and use
Inequality (\ref{eq:Beta_event_probability}).
We have
\begin{align*}
\nonumber
\sum_{j \in V \setminus X} \Pr & \left[ A_{i,j}^\ell(X)\right] 
- \sum_{j \in V \setminus X } \Pr \left[ A_{i,j}^\ell(X \cup \{u\})\right] \\
= & \sum_{j \in V \setminus X} \Pr \left[ B_{i,j}^\ell(X, u)\right]  \\
\ge & \sum_{j \in V \setminus Y} \Pr \left[ B_{i,j}^\ell(Y, u)\right] \\
= & \sum_{j \in V \setminus Y} \Pr \left[ A_{i,j}^\ell(Y)\right] 
- \sum_{j \in V \setminus Y } \Pr \left[ A_{i,j}^\ell(Y \cup \{u\})\right],
\end{align*} 
which completes the proof.
\end{IEEEproof}

\smallskip
Finally, we establish the hardness of $k$ absorbing centrality, 
defined in Problem~\ref{problem:k-ac}.

\begin{theorem}
The \krwc\ problem is \NP-hard.
\end{theorem}
\begin{IEEEproof}
We obtain a reduction from the {\VC} problem~\cite{GJ}.
An instance of the \VC\ problem is specified by a graph
$G = (V, E)$ and an integer $k$, and asks whether there exists a
set of nodes $C\subseteq V$ such that $|C|\le k$ and $C$ is a
vertex cover, 
(i.e., for every $(i,j) \in E$ it is $\{i,j\} \cap C\ne\emptyset$).
Let $|V|=n$.

Given an instance of the {\VC} problem, we construct an instance of
the decision version of {\krwc} by taking the same graph $G=(V,E)$ with
query nodes $Q=V$ and asking whether there is a set of absorbing nodes
$C$ such that $|C| \le k$ and 
$\abscentrality_{_Q}(C) \le 1 - \frac{k}{n}$.

We will show that $C$ is a solution for {\VC} 
if and only if $\abscentrality_{_Q}(C) \le 1 - \frac{k}{n}$. 

Assuming first that $C$ is a vertex cover. 
Consider a random walk starting uniformly at random from a node $v\in Q=V$.
If $v\in C$ then the length of the walk will be 0, 
as the walk will be absorbed immediately. 
This happens with probability $|C|/|V|=k/n$.
Otherwise, if  $v\not\in C$ the length of the walk will be 1, 
as the walk will be absorbed in the next step
(since $C$ is a vertex cover all the neighbors of $v$ need to
belong in $C$). 
This happens with the rest of the probability $1-k/n$.
Thus, the expected length of the random walk is
\begin{equation}
\abscentrality_{_Q}(C)  
=  0\cdot \frac{k}{n} + 1\cdot\left(1 - \frac{k}{n}\right) 
= 1 - \frac{k}{n} \\
\end{equation}
Conversely, assume that $C$ is not a vertex cover for $G$.
Then, there should be an uncovered edge $(u,v)$. 
A random walk that starts in $u$ and then goes to $v$ 
(or starts in $v$ and then goes to $u$)
will have length at least 2, 
and this happens with probability at least 
$\frac{2}{n}\frac{1}{d_{\max}} \ge \frac{2}{n^2}$. 
Then, following a similar reasoning as in the previous case, we have
\begin{eqnarray}
\nonumber
\abscentrality_{_Q}(C) &= & 
\sum_{k=0}^\infty k\, \Pr  \left( \text{absorbed in exactly } k \text{ steps}\right) \\ 
\nonumber
& =  & \sum_{k=1}^\infty \Pr \left( \text{absorbed after at least } k \text{ steps}\right) \\ 
& \ge & \left(1 - \frac{k}{n}\right) +  \frac{2}{n^2} > 1 - \frac{k}{n}.  
\end{eqnarray}
\end{IEEEproof}

\section{Algorithms}
\label{section:algorithms}

This section presents algorithms to solve the \krwc\ problem.
In all cases, the set of query nodes $Q\subseteq V$ is given as input,
along with a set of candidate nodes $D\subseteq V$ and the restart probability
$\alpha$.

\subsection{Greedy approach}
\label{section:k-ac-algorithms}

The first algorithm is a standard greedy algorithm, denoted \greedy, 
which exploits the
supermodularity of the absorbing random-walk centrality
measure. It starts with the result set $C$ equal to the empty set,
and iteratively adds a node from the set of candidate nodes $D$, 
until $k$ nodes are added.
In each iteration the node added in the set $C$ is the one that brings
the largest improvement to $\abscentrality_Q$.

As shown before, the objective function to be
minimized, i.e., $\abscentrality_Q$, is supermodular and monotonically decreasing.
The \greedy\ algorithm is not an approximation
algorithm for this minimization problem. However, it can be shown to
provide an approximation guarantee for maximizing the  
{\em absorbing centrality gain} measure, defined below.
\begin{definition}[Absorbing centrality gain]
Given a graph $G$,  
a set of query nodes $Q$, 
and a set of candidate nodes $D$, 
the absorbing centrality gain of a set of nodes 
$C\subseteq D$ is defined as 
\[
\acg_Q(C) = \m_Q - \abscentrality_Q(C),
\]
where
$\m_Q = {\min}_{v\in D}\{\abscentrality_Q(\{v\})\}$.
\end{definition}

\para{Justification of the gain function.}
The reason to define the absorbing centrality gain is
to turn our problem into a submodular-maxi\-mization problem 
so that we can apply standard approximation-theory results
and show that the greedy algorithm provides a constant-factor approximation guarantee.
The \emph{shift} $\m_Q$ quantifies the absorbing centrality of the best
single node in the candidate set.
Thus, the value of $\acg_Q(C)$
expresses how much we gain in expected random-walk length 
when we use the set $C$ as absorbing nodes 
compared to when we use the best single node.
Our goal is to maximize this gain. 

Observe that the gain function $\acg_Q$ is not non-negative everywhere. 
Take for example any node $u$
such that $\abscentrality_Q(\{u\}) > \m_Q$. Then, 
$\acg_Q(\{u\}) < 0$. Note also that we could have obtained a
non-negative gain function 
by defining gain with respect to the \emph{worst}
single node, instead of the best. 
In other words, the gain function
$\acg'_Q(C) = \M_Q - \abscentrality_Q(C)$,
with 
$\M_Q = {\max}_{v\in D}\{\abscentrality_Q(\{v\})\}$, 
is non-negative everywhere.

Nevertheless, the reason we use the gain function $\acg_Q$ 
instead of $\acg'_Q$ is that $\acg'_Q$ takes much larger
values than $\acg_Q$, and thus, a multiplicative approximation
guarantee on $\acg'_Q$ is a 
weaker result than a multiplicative approximation guarantee on $\acg_Q$.
On the other hand, 
our definition of $\acg_Q$ creates a technical difficulty with 
the approximation guarantee, 
that is defined for non-negative functions.
Luckily, this difficulty can be overcome easily
by noting that, due to the monotonicity of $\acg_Q$,
for any $k>1$, 
the optimal solution of the function $\acg_Q$, 
as well as the solution returned by \greedy, 
are both non-negative. 

\para{Approximation guarantee.}
The fact that the \greedy\ algorithm gives an approximation guarantee 
to the problem of maximizing absorbing centrality gain is a standard 
result from the theory of submodular functions.

\begin{proposition}
The function $\acg_Q$ is monotonically increasing, and
submodular.
\end{proposition}

\begin{proposition}
Let $k>1$.
For the problem of finding a set $C\subseteq D$ with $|C|\le k$, 
such that $\acg_Q(C)$ is maximized, the \greedy\ algorithm gives a
$\left(1-\frac{1}{e}\right)$-approximation guarantee.
\end{proposition}

We now discuss the complexity of the \greedy\ algorithm.
A na\"ive implementation 
requires computing the absorbing
centrality $\abscentrality_Q(C)$ using Equation~(\ref{eq:inversion})
for each set $C$ that needs to be evaluated during the execution of
the algorithm.
However, applying Equation~(\ref{eq:inversion}) involves a matrix
inversion, which is a very expensive operation.
Furthermore, the number of times that we need to evaluate
$\abscentrality_Q(C)$ is ${\cal O}(k|D|)$, as for each iteration of
the greedy we need to evaluate the improvement over the current set of
each of the ${\cal O}(|D|)$ candidates.
The number of candidates can be very large, e.g., $|D|=n$, yielding an 
${\cal O}(kn^4)$ algorithm, which is prohibitively expensive.

We can show, however, that we can execute \greedy\ significantly more
efficiently. Specifically, we can prove the following two propositions.
\begin{proposition}
Let $C_{i-1}$ be a set of $i-1$ absorbing nodes, 
$\mathbf{P}_{i-1}$ the corresponding
transition matrix, and let $\mathbf{F}_{i-1} = (\mathbf{I}-\mathbf{P}_{i-1})^{-1}$.
Let $C_i = C_{i-1} \cup \{u\}$. 
Given  $\mathbf{F}_{i-1}$ the value $\abscentrality_Q(C_i)$ can be computed in $\mathcal{O}(n^2)$.
\label{prop:fast-updates-next-step}
\end{proposition}
\begin{proposition}
Let $C$ be a set of absorbing nodes, $\mathbf{P}$ the corresponding
transition matrix, and $\mathbf{F} = (\mathbf{I}-\mathbf{P})^{-1}$.
Let $C' = C - \{v\}  \cup \{u\}$, $u, v \in C$. 
Given  $\mathbf{F}$ the value $\abscentrality_Q(C')$ can be computed in time $\mathcal{O}(n^2)$.
\label{prop:fast-updates-same-step}
\end{proposition}
The proofs of these two propositions can be found in the Appendix.
Proposition~\ref{prop:fast-updates-next-step} implies that in order to compute~$\abscentrality_Q(C_i)$
for absorbing nodes $C_i$ in $\mathcal{O}(n^2)$, it is enough to maintain the matrix $\mathbf{F}_{i-1}$,
computed in the previous step of the greedy algorithm for absorbing nodes $C_{i-1}$.
Proposition~\ref{prop:fast-updates-same-step}, on the other hand, implies that we can compute
the \absorbingcentrality\ of each set of absorbing nodes of a fixed size~$i$ in $\mathcal{O}(n^2)$,
given the matrix $\mathbf{F}$, which is computed for one arbitrary set of absorbing nodes $C$
of size $i$. Combined, the two propositions above yield a greedy algorithm that runs in
${\cal O}(kn^3)$ and offers the approximation guarantee discussed above. We outline it as
Algorithm~\ref{algo:greedy}.

\begin{algorithm}[t]
\caption{\greedy}
\label{algo:greedy}
\begin{algorithmic}
\STATE {\bf Input}: graph $G$, query nodes $Q$, candidates $D$, $k\geq 1$
\STATE {\bf Output}: a set of $k$ nodes $C$
\STATE Compute $\abscentrality_Q(\{v\})$ for arbitrary $v\in D$
\STATE For each $u\in (D - \{v\})$, use Prop.\ref{prop:fast-updates-same-step} to compute $\abscentrality_Q({u})$
\STATE Select $u_1\in D$ s.t.\ $u_1 \leftarrow \arg\max_{u\in D} \abscentrality_Q({u})$
\STATE Initialize solution $C \leftarrow \{u_1\}$
\FOR{$i = 2 .. k$}
\STATE For each $u\in D$, use Prop.\ref{prop:fast-updates-next-step} to compute $\abscentrality_Q(C\cup\{u\})$
\STATE Select $u_i \in D$ s.t.\ $u_i \leftarrow \arg\max_{u_i\in (D-C)} \abscentrality_Q(C\cup\{u\})$
\STATE Update solution $C \leftarrow C \cup \{u_i\}$
\ENDFOR
\STATE \textbf{return} $C$
\end{algorithmic}
\end{algorithm}

\para{Practical speed-up.} 
We  found that the following heuristic lets us speed-up \greedy\ even further, 
with no significant loss in the quality of results. 
To select the first node for the solution set $C$ (see Algorithm~\ref{algo:greedy}), 
we calculate the {\it PageRank} values of all nodes in $D$ and evaluate $\abscentrality_Q$ 
only for the $t << k$ nodes with highest PageRank score, 
where $t$ is a fixed parameter. 
In what follows, we will be using this heuristic version of \greedy, unless explicitly stated otherwise.

\subsection{Efficient heuristics}

Even though \greedy\ runs in polynomial time, it can
be quite inefficient when employed on moderately sized
datasets (more than some tens of thousands of nodes).
We thus describe algorithms that we study as
efficient heuristics for the problem.
These algorithms do not offer guarantee for their performance.

\spara{Spectral}
methods have been used extensively for the problem of
graph partitioning.
Motivated by the wide applicability of this family of algorithms,
here we explore three spectral algorithms: 
\spectralqueries, \spectralcandidates, and
\spectraldistance.
We start by a brief overview of the spectral method; 
a comprehensive presentation can be found in the 
tutorial by von Luxburg~\cite{von2007tutorial}.
 
The main idea of spectral approaches is to project the original graph
into a low-dimensional Euclidean space so that distances between
nodes in the graph correspond to Euclidean distances between the
corresponding projected points. 
A standard spectral embedding method, proposed by Shi and
Malik~\cite{Shi:2000gf},
uses the ``random-walk'' {\em Laplacian} matrix
$\mathbf{L}_G=\mathbf{I}-\mathbf{D}^{-1}\mathbf{A}$
of a graph $G$, where $\mathbf{A}$ is the adjacency matrix of the graph, 
and forms the matrix 
$\mathbf{U} =[u_2,\ldots,u_{d+1}]$ 
whose columns are the 
eigenvectors of  $\mathbf{L}_G$
that correspond to the smallest eigenvalues $\lambda_2\le\ldots\le\lambda_{d+1}$, 
with $d$ being the target dimension of the projection.
The spectral embedding is then defined by mapping 
the $i$-th node of the graph to a point in~$\mathbb{R}^d$, 
which is the $i$-row of the matrix~$\mathbf{U}$. 

The algorithms we explore are adaptations of the spectral method.
They all start by computing the spectral embedding $\phi:V\rightarrow\mathbb{R}^d$, 
as described above, and then, proceed as follows:

\sparanbf{\spectralqueries} 
performs $k$-means clustering on the embeddings
of the {\it query nodes}, where $k$ is the desired size of the result set.
Subsequently, it selects {\it candidate nodes} that are close to the
computed centroids.
Specifically, if $s_i$ is the size of the $i$-th cluster,
then $k_i$ candidate nodes are selected whose embedding is the nearest 
to the $i$-th centroid. The number $k_i$ is selected so that
$k_i \propto s_i$ and $\sum k_i = k$. 

\sparanbf{\spectralcandidates}
is similar to \spectralqueries, 
but it performs the  $k$-means clustering on the embeddings
of the {\it candidate nodes}, 
instead of the query nodes. 

\sparanbf{\spectraldistance}
performs $k$-means clustering on the embeddings
of the {\it query nodes}, where $k$ is the desired result-set size.
Then, it selects the $k$ candidate nodes whose embeddings minimize
the sum of squared $\ell_2$-distances from the centroids, 
with no consideration of the relative sizes of the clusters.

\spara{Personalized Pagerank} (\personalpagerank).
This is the standard Pagerank~\cite{PageRank} algorithm 
with a damping factor equal to the restart probability $\alpha$ of the random walk 
and personalization probabilities equal to the start probabilities  $\startprob(q)$. 
Algorithm \personalpagerank\ returns the $k$ nodes with highest PageRank values.

\spara{Degree and distance centrality.}
Finally, we consider the standard degree and distance centrality measures.

\sparanbf{\degree} returns the $k$ highest-degree nodes.
Note that this baseline is oblivious to the query nodes.

\sparanbf{\closest} returns the $k$ nodes with highest distance
centrality with respect to $Q$. 
The distance centrality
of a node $u$
is defined as 
$\distcentrality(u) = \left( \sum_{v\in Q} d(u,v)\right)^{-1}$.

\section{Experimental evaluation}
\label{sec:experiments}

\subsection{Datasets}
\label{section:datasets}

We evaluate the algorithms described in Section~\ref{section:algorithms}
on two sets of real graphs:
one set of small graphs that allows us to compare the performance of
the fast heuristics against the greedy approach; 
and one set of larger graphs, 
to compare the performance of the heuristics against
each other on datasets of larger scale. 
Note that the bottleneck of the computation lies in the evaluation
of centrality. 
Even though the technique we describe in
Section~\ref{section:fasteval} allows it to scale 
to datasets of tens of thousands of nodes on a single processor,
it is still prohibitively expensive for massive graphs.
Still, our experimentation allows us to discover the traits of the different
algorithms and understand what performance to anticipate when they are
employed on graphs of massive size.

The datasets are listed in Table~\ref{table:datasets}. Small graphs are obtained from
Mark Newman's repository\footnote{{http://www-personal.umich.edu/\%7Emejn/netdata/}}, 
larger graphs from SNAP.\footnote{\url{http://snap.stanford.edu/data/index.html}}
For \dblpcoauthors, \livejournal, and \roadnet\
we use samples of the original datasets. 
In the interest of repeatability, our code and datasets are
made publicly available.\footnote{{https://github.com/harrymvr/absorbing-centrality}}

\begin{table}[t]
\centering
\caption{Dataset statistics}
\begin{tabular}{lrr}
\toprule
Dataset   & $|V|$ & $|E|$ \\
\midrule
\karate & $34$ & $78$ \\
\dolphins & $62$ & $159$ \\
\lesmis & $77$ & $254$ \\
\adjnoun & $112$ & $425$ \\
\football & $115$ & $613$ \\ \hline
\dblpcoauthors & $2\,891$   & $2\,891$\\
\livejournal & $3\,645$ & $4\,141$ \\
\cagrqc & $5\,242$ & $14\,496$\\
\cahepth & $9\,877$ & $25\,998$\\
\roadnet & $10\,199$ & $13\,932$\\
\oregon & $11\,174$ & $23\,409$\\
\bottomrule
\end{tabular}
\label{table:datasets}
\end{table}


\subsection{Evaluation Methodology}

Each experiment in our
evaluation framework is defined by 
a graph~$G$, 
a set of query nodes~$Q$, 
a set of candidate nodes~$D$, and
an algorithm to solve the problem.
We evaluate all algorithms presented in Section~\ref{section:algorithms}.
For the set of candidate nodes~$D$, we consider two cases: 
it is equal to either the set of query nodes, i.e., $D = Q$, 
or the set of all nodes, i.e., $D = V$.

Query nodes $Q$ are selected randomly, 
using the following process:
First, we select a set $S$ of $\nsph$ seed nodes, uniformly at random
among all nodes. 
Then, we select a ball $B(v,r)$ 
of predetermined radius $r = 2$,
around each seed $v\in S$.\footnote{For the planar
\roadnet\ dataset we use $r = 3$.}
Finally, from all balls, 
we select a set of query nodes $Q$ of predetermined size $q$, 
with $q = 10$ and $q = 20$,
respectively, for the small and larger datasets. 
Selection is done uniformly at random. 

Finally, 
the restart probability $\alpha$ is set to $\alpha=0.15$ 
and the starting probabilities $\startprob$ are uniform over $Q$.
 
\subsection{Implementation}
All algorithms are implemented in Python using the 
NetworkX package~\cite{hagberg-2008-exploring}, 
and were run on an Intel Xeon 2.83GHz with 32GB RAM.

\subsection{Results}

\begin{figure*}[t]
        \centering
        \begin{subfigure}[b]{0.49\textwidth}
                \includegraphics[width=\textwidth]{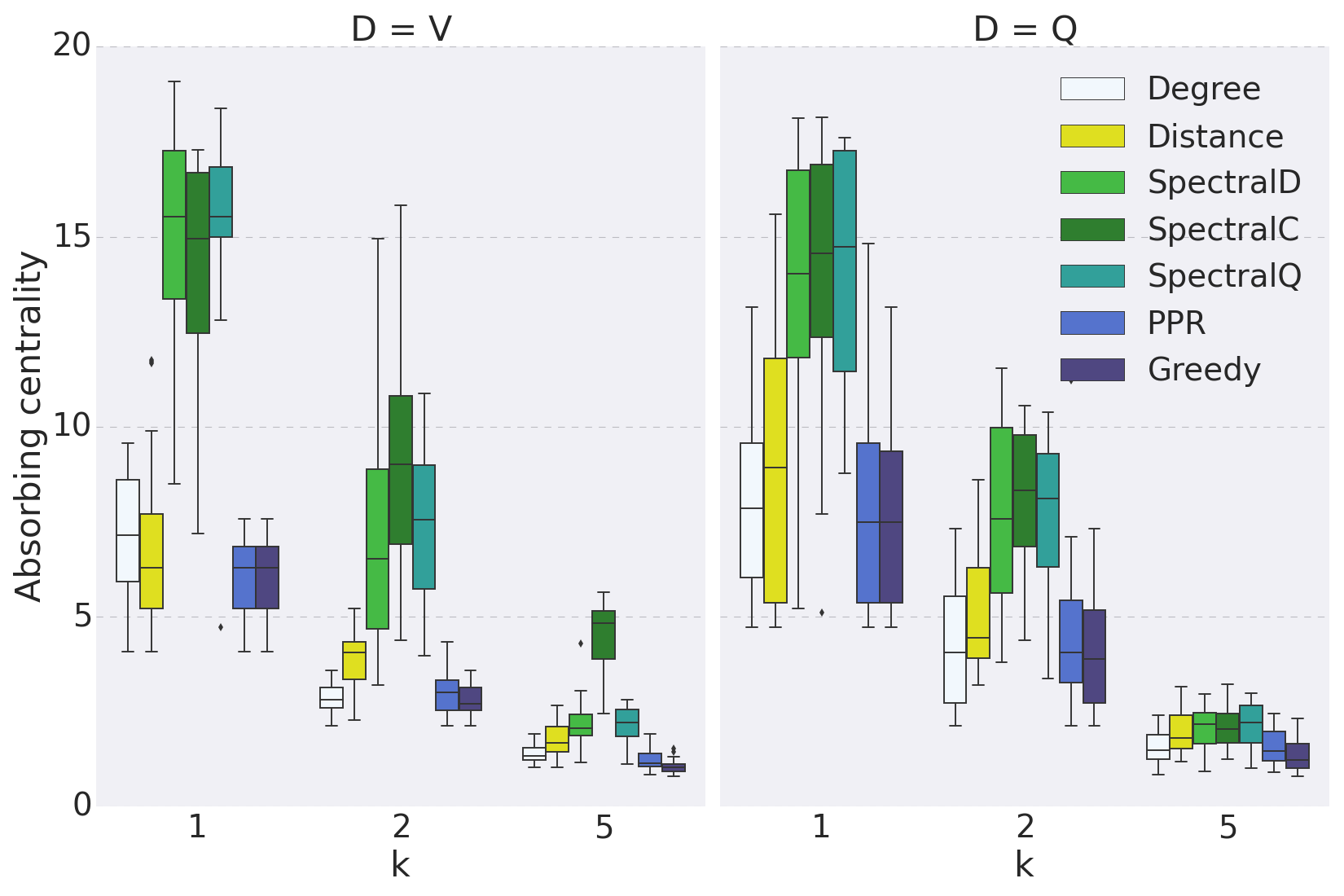}
                \caption{\karate}
        \end{subfigure}%
        ~ 
        \begin{subfigure}[b]{0.49\textwidth}
                \includegraphics[width=\textwidth]{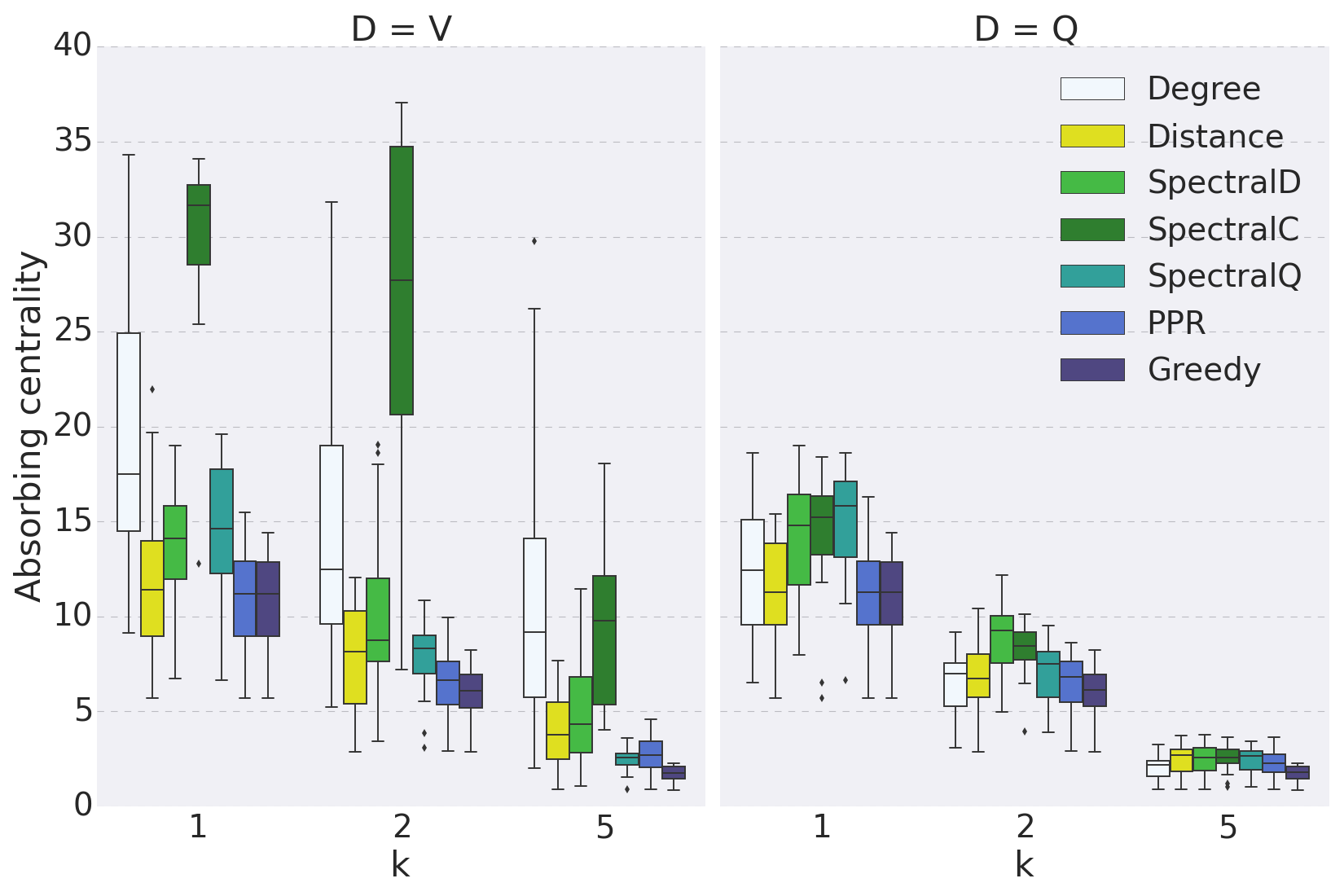}
                \caption{\dolphins}
        \end{subfigure}

        \begin{subfigure}[b]{0.49\textwidth}
                \includegraphics[width=\textwidth]{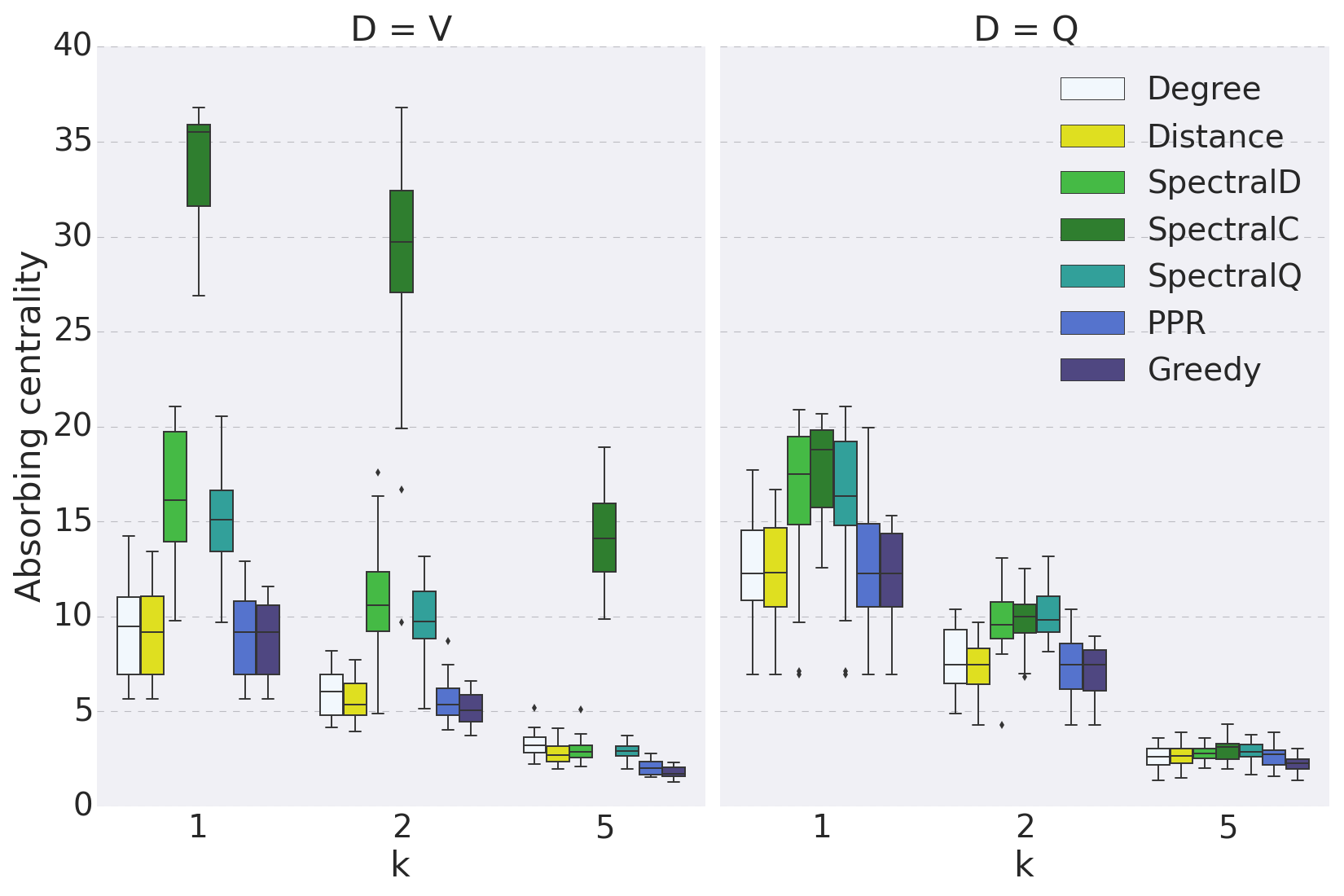}
                \caption{\lesmis}
        \end{subfigure}
        ~
        \begin{subfigure}[b]{0.49\textwidth}
                \includegraphics[width=\textwidth]{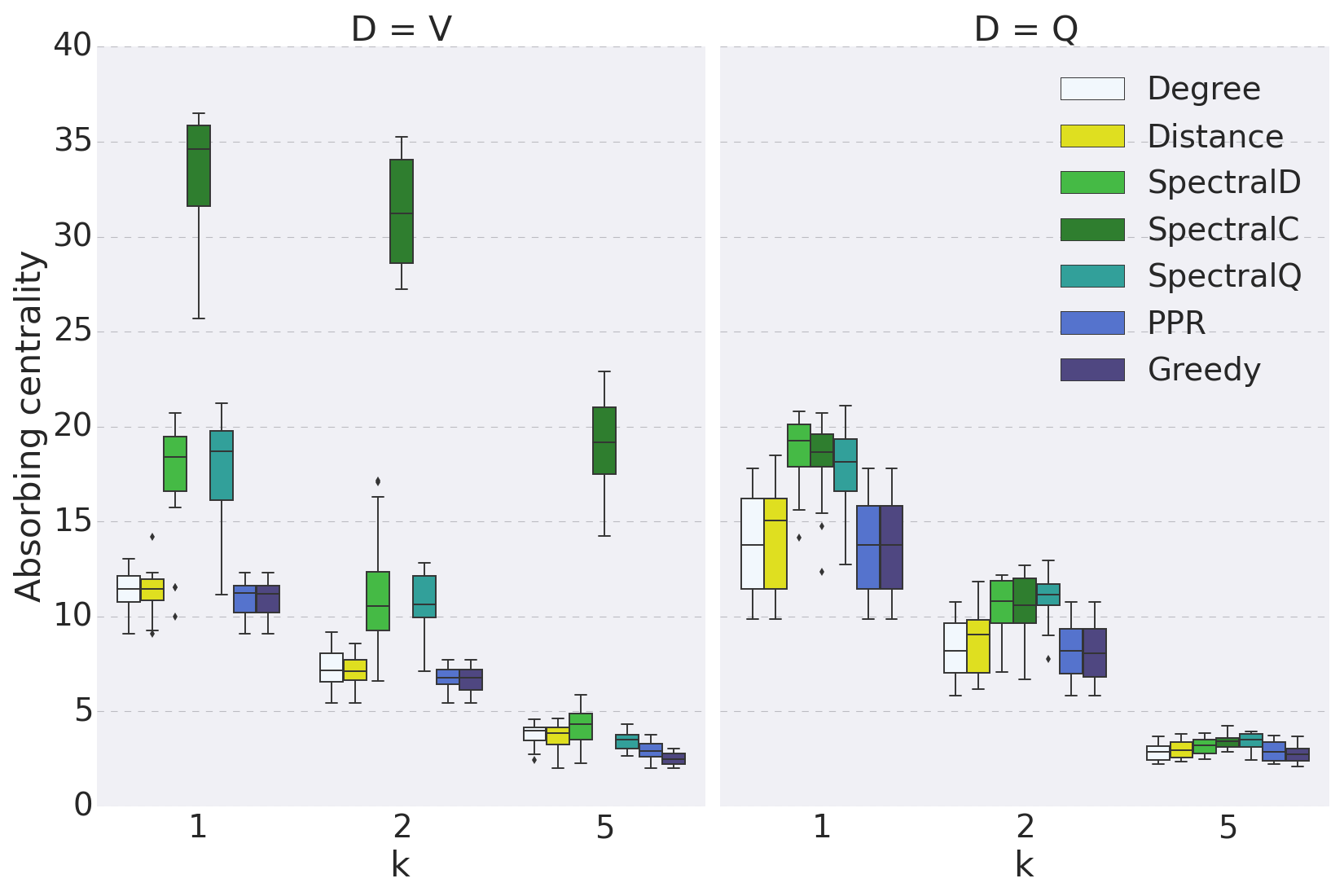}
                \caption{\adjnoun}
        \end{subfigure}

        \begin{subfigure}[b]{0.49\textwidth}
                \includegraphics[width=\textwidth]{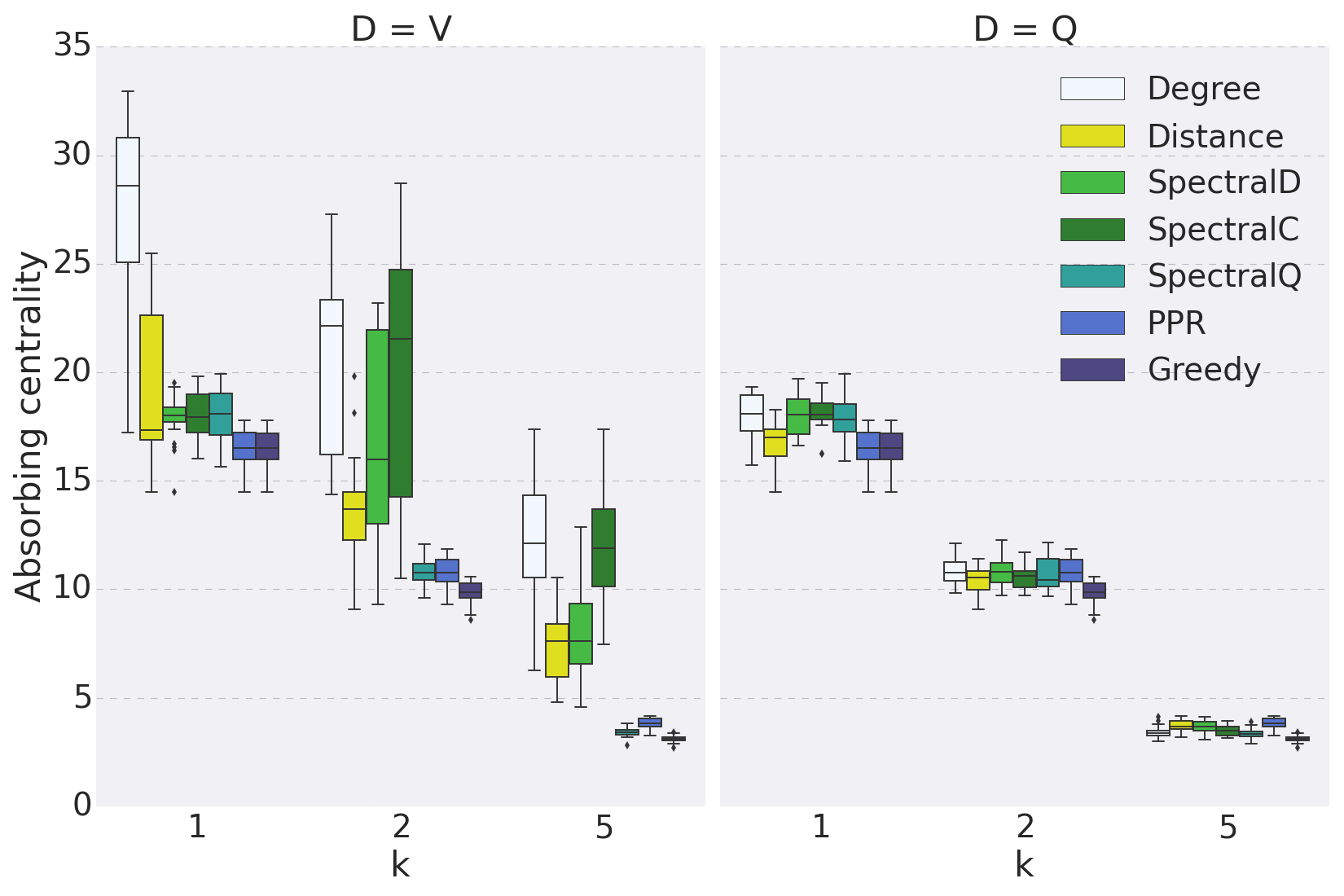}
                \caption{\football}
        \end{subfigure}
        \caption{Results on small datasets for varying $k$ and $\nsph = 2$.}
        \label{fig:small}
\end{figure*}

\begin{figure*}[h]
        \centering
        \begin{subfigure}[b]{0.49\textwidth}
                \includegraphics[width=\textwidth]{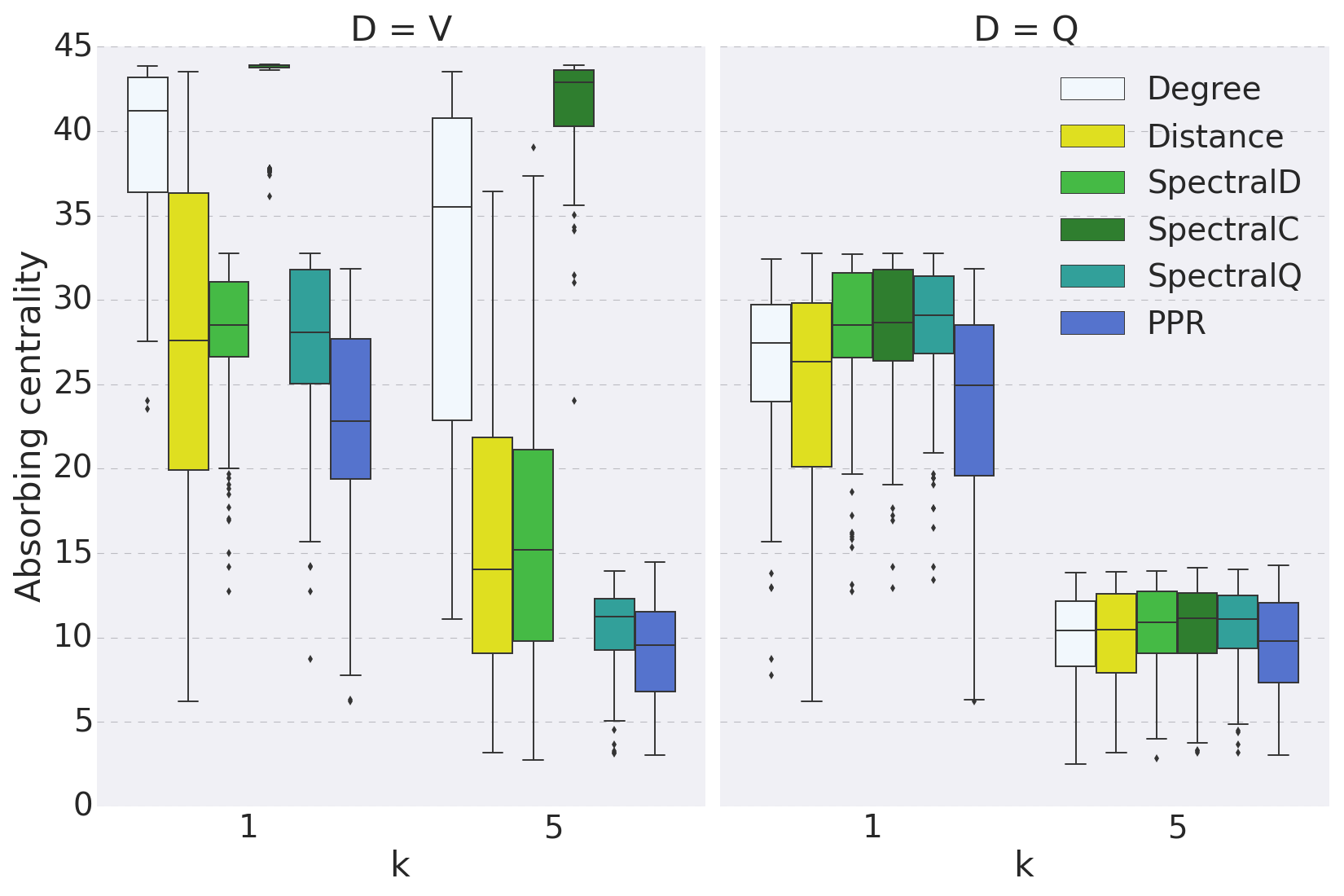}
                \caption{\cagrqc}
        \end{subfigure}%
        ~
        \begin{subfigure}[b]{0.49\textwidth}
                \includegraphics[width=\textwidth]{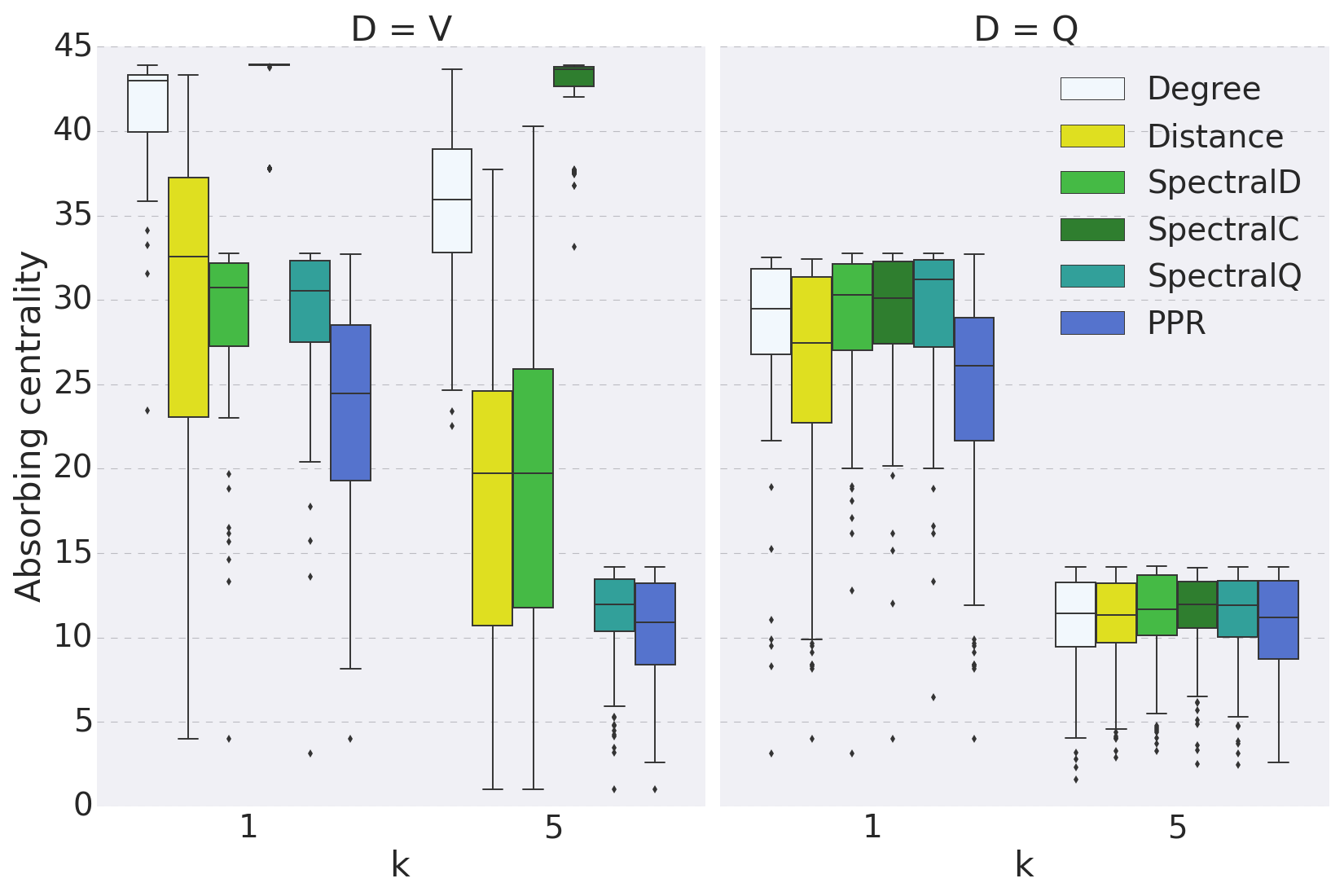}
                \caption{\cahepth}
        \end{subfigure}
        
        \begin{subfigure}[b]{0.49\textwidth}
                \includegraphics[width=\textwidth]{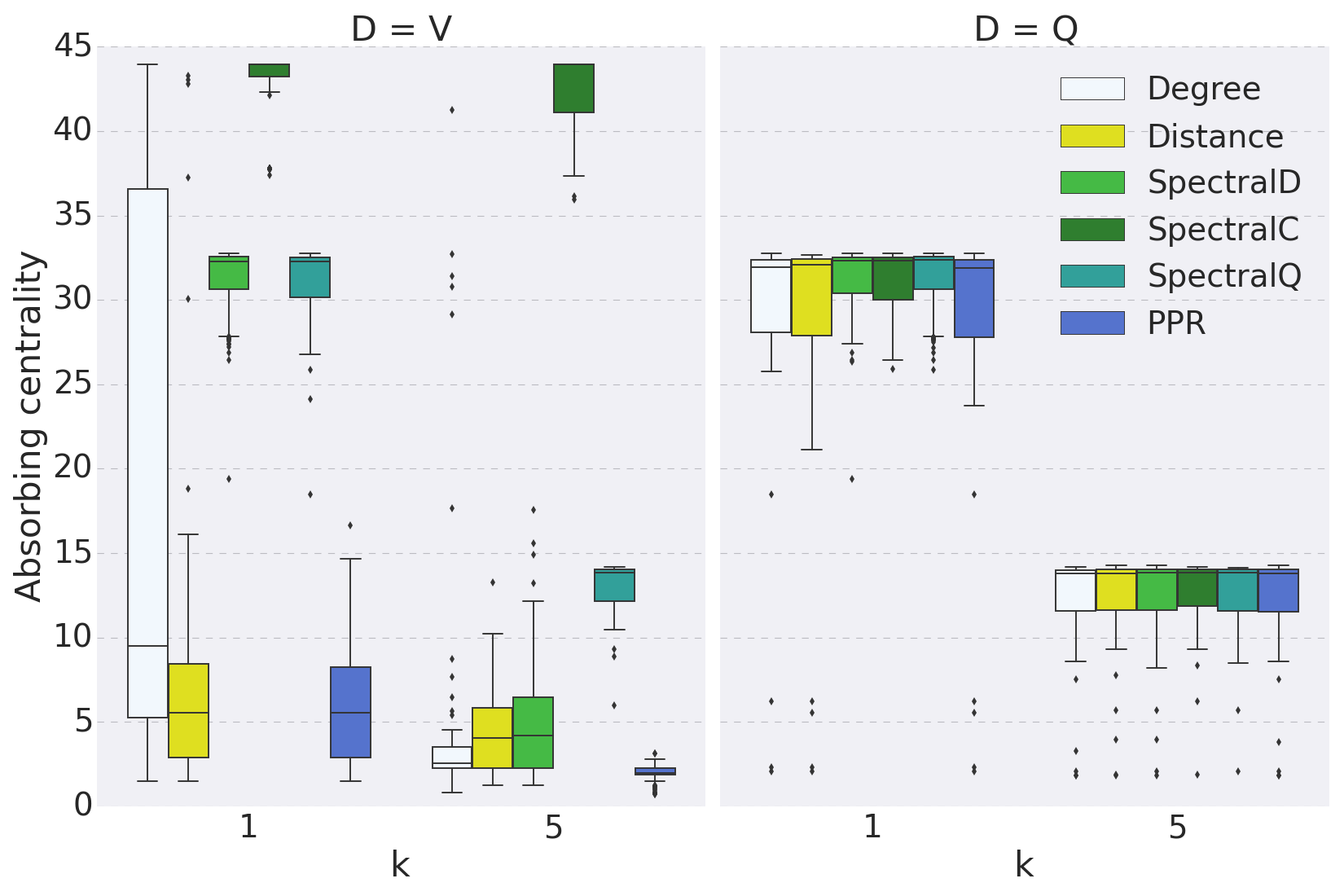}
                \caption{\livejournal}
        \end{subfigure}
        ~
        \begin{subfigure}[b]{0.49\textwidth}
                \includegraphics[width=\textwidth]{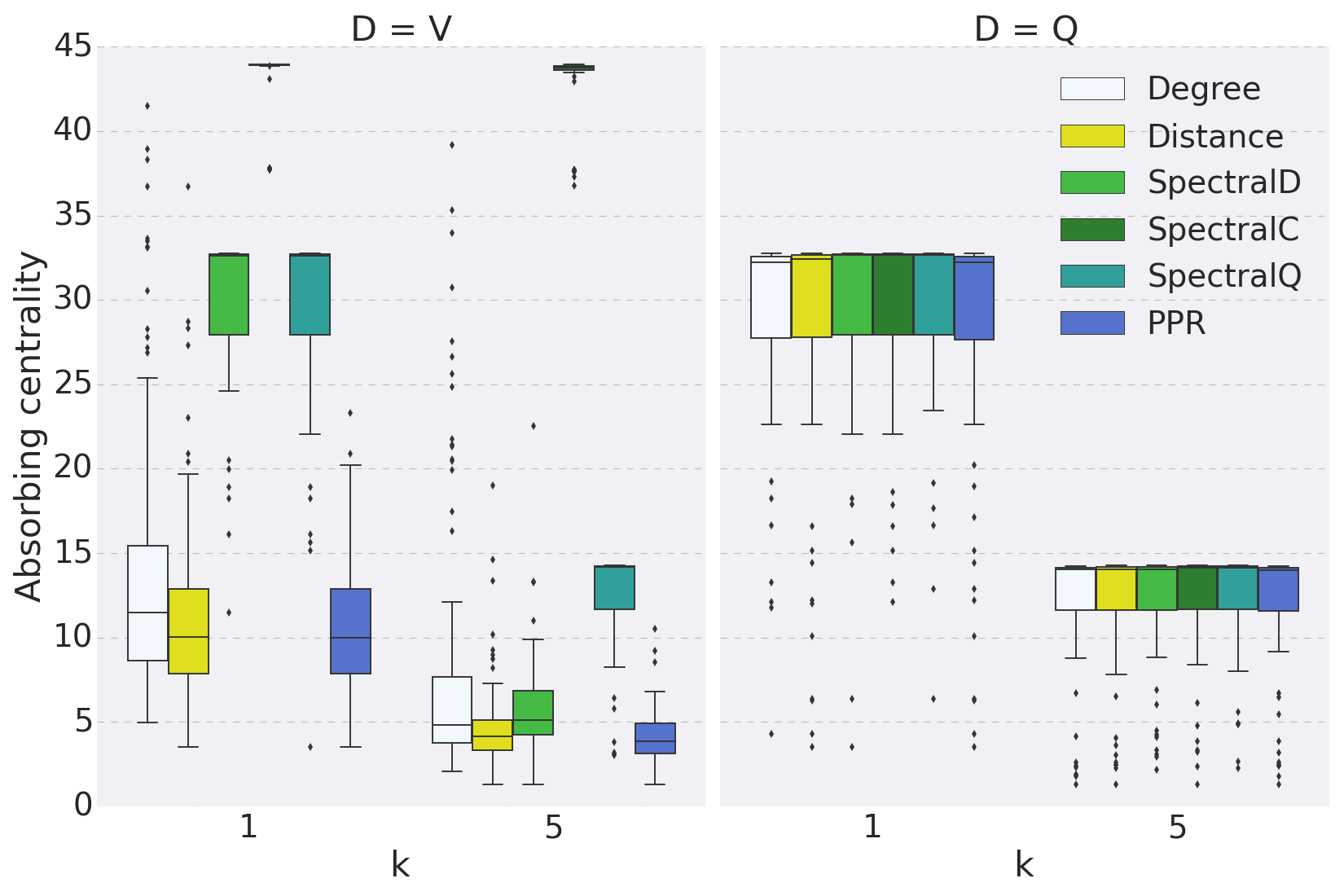}
                \caption{\oregon}
        \end{subfigure}
        
        \begin{subfigure}[b]{0.49\textwidth}
                \includegraphics[width=\textwidth]{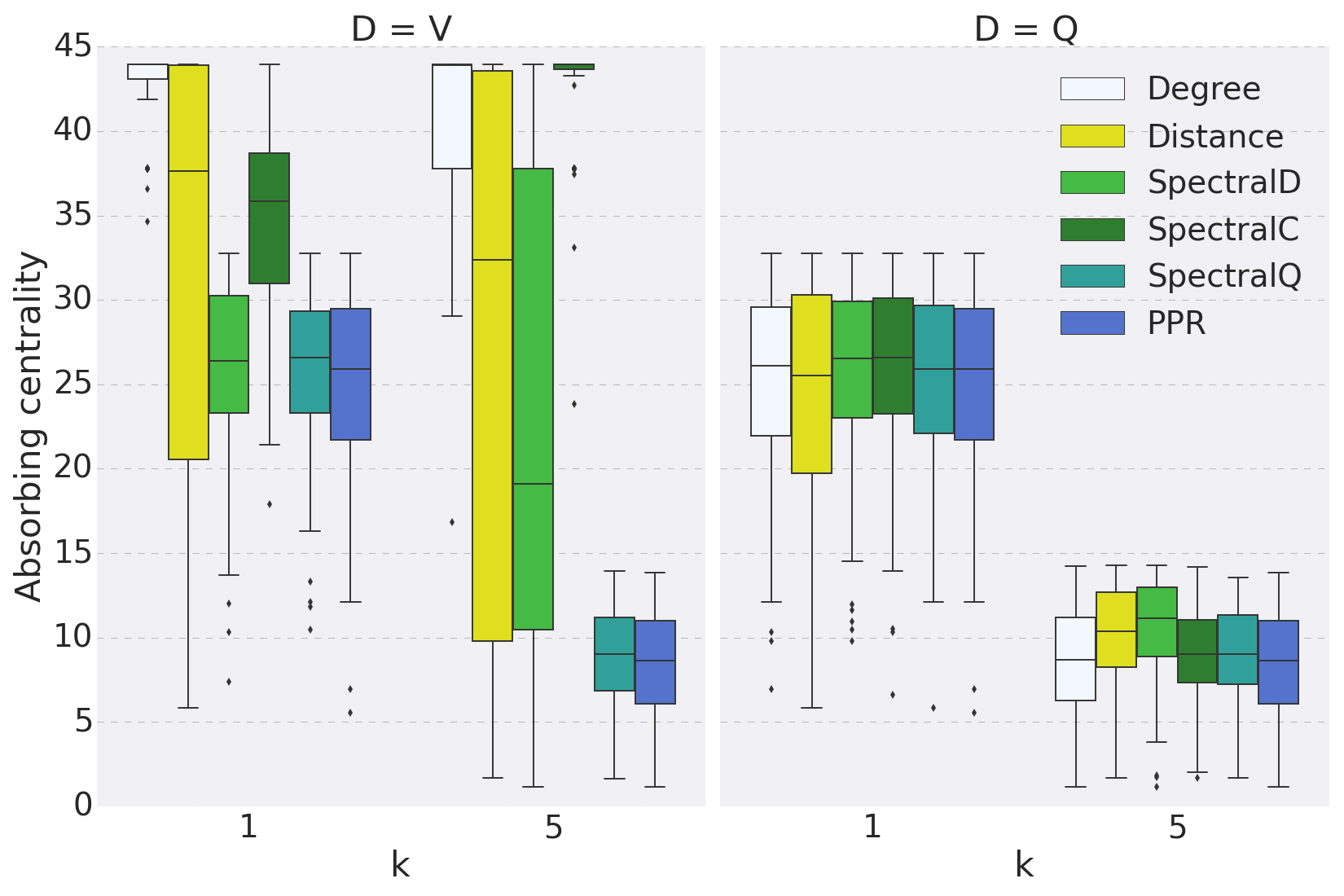}
                \caption{\roadnet}
        \end{subfigure}
        ~
        \begin{subfigure}[b]{0.49\textwidth}
                \includegraphics[width=\textwidth]{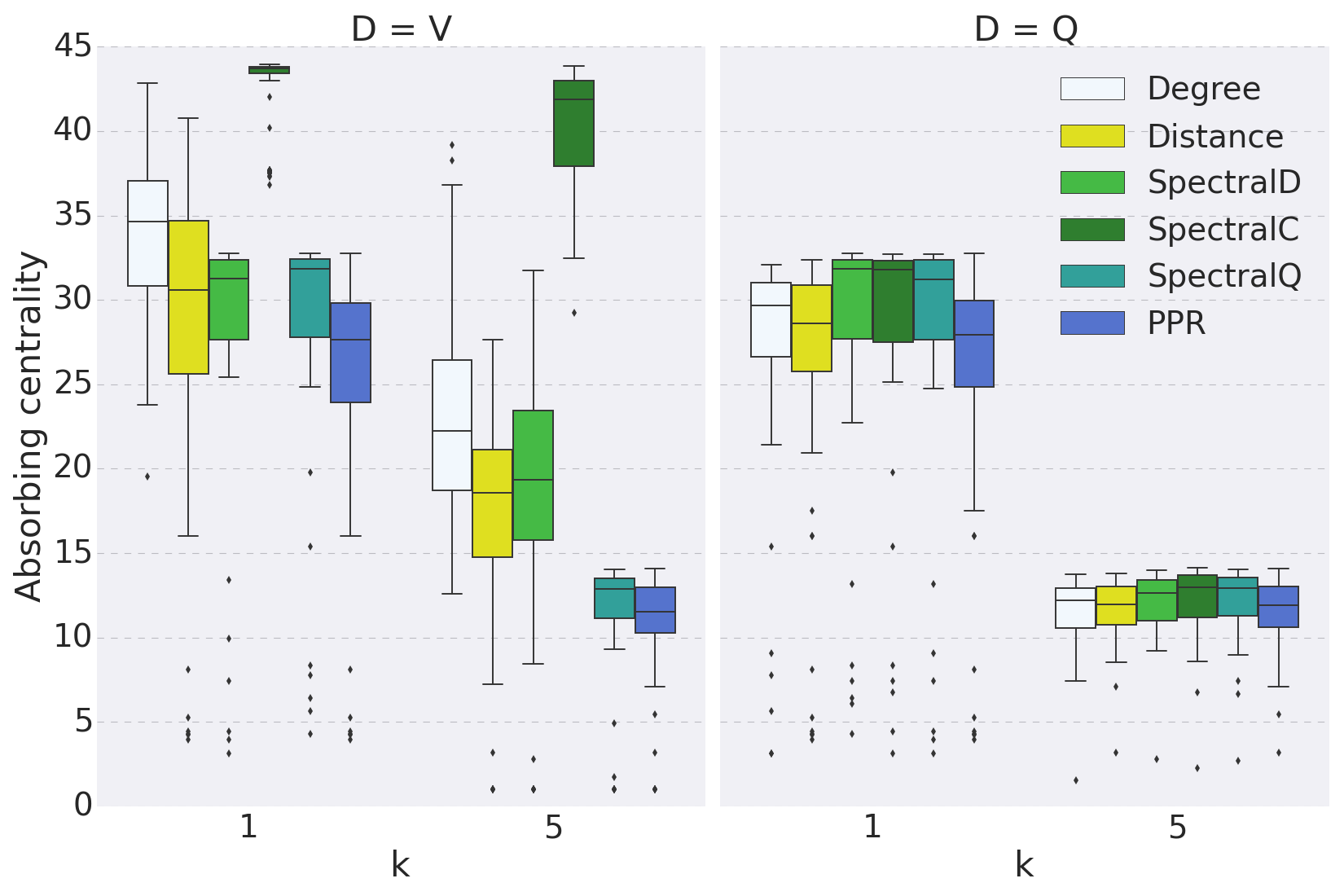}
                \caption{\dblpcoauthors}
        \end{subfigure}
        \caption{Results on large datasets for varying $k$ and $\nsph = 5$.}
        \label{fig:large_quality}
\end{figure*}

Figure~\ref{fig:small} shows the centrality scores achieved by different algorithms 
on the small graphs for varying $k$ (note: lower is better).
We present two settings: on the left, the candidates are all nodes ($D = V$), 
and on the right, the candidates are only the query nodes ($D = Q$).
We observe that \personalpagerank\ tracks well
the quality of solutions returned by \greedy, 
while \degree\ and \closest\ often come close to that. 
Spectral algorithms do not perform that well.

Figure~\ref{fig:large_quality} is similar to Figure~\ref{fig:small}, but
results on the larger datasets are shown, not including \greedy. 
When all nodes are candidates, 
\personalpagerank\ typically has the best performance,  
followed by \closest, 
while \degree\ is unreliable. 
The spectral algorithms typically perform worse than \personalpagerank.

When only query nodes are candidates,
all algorithms demonstrate similar performance, 
which is most typically worse than the performance of
\personalpagerank\ (the best performing algorithm) in the previous setting. 
Both observations can be explained by the fact that  
the selection is very restricted by the requirement $D = Q$, 
and there is not much flexibility for the best
performing algorithms 
to produce a better solution.

In terms of running time on the larger graphs, \closest\ returns
within a few minutes 
(with observed times between 15 seconds to 5 minutes)
while \degree\ returns within seconds 
(all observed times were less than 1 minute). 
Finally, even though \greedy\ returns within 1-2 seconds for the small datasets, 
it does not scale well for the larger datasets 
(running time is orders of magnitude worse than the heuristics and not included in the experiments).

Based on the above, we conclude that \personalpagerank\ offers the best trade-off of quality 
versus running time for datasets of at least moderate size (more than $10\,\text{k}$ nodes).

\section{Conclusions}
\label{sec:conclusions}

In this paper, we have addressed the problem of 
finding central nodes in a graph with respect to a set of query nodes~$Q$.
Our measure is based on absorbing random walks:
we seek to compute $k$ nodes that minimize 
the expected number of steps that a random walk will need to
reach at (and be ``absorbed'' by) when it starts from the query nodes.
We have shown that the problem is \NP-hard and
described an $\mathcal{O}(kn^3)$ greedy algorithm to solve it approximately.
Moreover, we experimented with heuristic algorithms to solve the problem on
large graphs. 
Our results show that, in practice, personalized PageRank
offers a good combination of quality and speed.

\bibliographystyle{abbrv}
\bibliography{bibliography}

\appendix


\subsection{Proposition~\ref{proposition:monotonicity}}

\begin{proposition-non}[Monotonicity]
For all $X\subseteq Y\subseteq V$ it is
$\abscentrality(Y)\le\abscentrality(X)$.
\end{proposition-non}
\begin{IEEEproof}
Write $G_X$ for the input graph $G$ where the set $X$ are absorbing
nodes.
Define $G_Y$ similarly.
Let $Z=Y\setminus X$.
Consider a path $p$ in $G_X$ drawn from the distribution induced by
the random walks on $G_X$.
Let $\Pr\left[p\right]$ be the probability of the path and $\ell(p)$ its length.
Let ${\cal P}(X)$ and ${\cal P}(Y)$  
be the set of paths on $G_X$ and $G_Y$.
Finally, 
let ${\cal P}(Z, X)$ be the set of paths on $G_X$ 
that pass from $Z$, and 
${\cal P}(\overline{Z}, X)$  the set of paths on $G_X$ 
that do {\em not} pass from $Z$.
We have
\begin{eqnarray*}
\abscentrality(X) 
& = & \sum_{p\in{\cal P}(X)} \Pr\left[p\right]\ell(p) \\
& = & \sum_{p\in{\cal P}(\overline{Z},X)} \Pr\left[p\right]\ell(p) +
\sum_{p\in{\cal P}(Z,X)} \Pr\left[p\right]\ell(p) \\
& \ge & \sum_{p\in{\cal P}(Y)} \Pr\left[p\right]\ell(p) \\
& = & \abscentrality(Y),
\end{eqnarray*}
where the inequality comes from the fact that a path 
in $G_X$ passing from $Z$ and being absorbed by $X$ corresponds to a
shorter path in $G_Y$ being absorbed by $Y$.
\end{IEEEproof}

\bigskip
\subsection{Proposition~\ref{prop:fast-updates-next-step}}
\label{appe:fast-updates-next-step}

\begin{proposition-non}
Let $C_{i-1}$ be a set of $i-1$ absorbing nodes, $\mathbf{P}_{i-1}$
the corresponding transition matrix, and
$\mathbf{F}_{i-1} = (\mathbf{I}-\mathbf{P}_{i-1})^{-1}$.
Let $C_i = C_{i-1} \cup \{u\}$. Given  $\mathbf{F}_{i-1}$,
the centrality score
$\abscentrality_Q(C_i)$ can be computed in time $\mathcal{O}(n^2)$.
\end{proposition-non}

The proof makes use of the following lemma.

\begin{lemma}[Sherman-Morrison Formula~\cite{golub2012matrix}]
Let $\mathbf{M}$ be a square $n \times n$ invertible matrix and $\mathbf{M}^{-1}$ its inverse.
Moreover, let $\mathbf{a}$ and $\mathbf{b}$ be any two column vectors of size $n$.
Then, the following equation holds
$$(\mathbf{M} + \mathbf{a}\mathbf{b}^{T})^{-1} = 
\mathbf{M}^{-1} - \mathbf{M}^{-1}\mathbf{a}\mathbf{b}^{T}\mathbf{M}^{-1} 
/ (1 + \mathbf{b}^{T}\mathbf{M}^{-1}\mathbf{a}).$$
\label{lemma:woodbury-simple}
\end{lemma}
\begin{IEEEproof}
Without loss of generality, let the set of absorbing nodes be
$C_{i-1} = \{1,2,\ldots,i-1\}$.
As in Section~\ref{section:algorithms}, the expected
number of steps before absorption is given by the formulas
\begin{equation*}
\abscentrality_{_Q}(C_{i-1}) =
\startprob^{T}_{_Q} \mathbf{F}_{i-1} \mathbf{1}, 
\end{equation*}
\begin{equation*}
\text{with } 
\mathbf{F}_{i-1} = \mathbf{A}_{i-1}^{-1}
\text{ and } 
\mathbf{A}_{i-1} = \mathbf{I} - \mathbf{P}_{i-1}.
\end{equation*}
We proceed to show how to increase the set of absorbing
nodes by one and calculate the new absorption time
by updating $\mathbf{F}_{i-1}$ in $\mathcal{O}(n^2)$.
Without loss of generality, suppose we add node $i$ to
the absorbing nodes $C_{i-1}$, so that
$$C_i = C_{i-1} \cup \{i\} = \{1,2,\ldots,i-1,i\}.$$
Let $\mathbf{P}_i$ be the transition matrix over $G$
with absorbing nodes $C_i$.
Like before, the expected absorption time by nodes $C_i$
is given by the formulas
\begin{equation*}
\abscentrality_{_Q}(C_{i}) =
\startprob^{T}_{_Q} \mathbf{F}_{i} \mathbf{1}, 
\end{equation*}
\begin{equation*}
\text{with }
\mathbf{F}_{i} = \mathbf{A}_{i}^{-1}
\text{ and }
\mathbf{A}_{i} = \mathbf{I} - \mathbf{P}_{i}.
\end{equation*}
Notice that 
\begin{eqnarray*}
& \mathbf{A}_i - \mathbf{A}_{i-1} = 
(\mathbf{I} - \mathbf{P}_i) - (\mathbf{I} - \mathbf{P}_{i-1}) =
\mathbf{P}_{i-1} - \mathbf{P}_i & \\
& =
\left[
\begin{array}{c}
\mathbf{0}_{(i-1)\times n} \\
p_{i, 1} \ldots p_{i, n} \\
\mathbf{0}_{(n-i)\times n}  
\end{array}
\right]
= \mathbf{a} \mathbf{b}^{T} &  
\end{eqnarray*}
where $p_{i, j}$ denotes the transition probability from node
$i$ to node $j$ in transition matrix $\mathbf{P}_{i-1}$, and
the column-vectors $\mathbf{a}$ and $\mathbf{b}$ 
are defined as
\begin{eqnarray*}
\mathbf{a} & = & [\overbrace{0 \ldots 0}^{i-1} \ 1 \ \overbrace{0 \ldots 0}^{n-i}], \,\text{ and}\\
\mathbf{b} & = & [p_{i,1} \ldots p_{i,n}].
\end{eqnarray*}
By a direct application of Lemma~\ref{lemma:woodbury-simple},
it is easy to see that we can compute $\mathbf{F}_i$ from $\mathbf{F}_{i-1}$
with the following formula, at a cost of $\mathcal{O}(n^2)$ operations.
\begin{eqnarray*}
\mathbf{F}_{i} & = & \mathbf{F}_{i-1} - (\mathbf{F}_{i-1} \mathbf{a})(\mathbf{b}^{T} \mathbf{F}_{i-1}) 
/ (1 + \mathbf{b}^{T}(\mathbf{F}_{i-1} \mathbf{a})) 
\end{eqnarray*}
We have thus shown that, given $\mathbf{F}_{i-1}$, we can compute $\mathbf{F}_i$,
and therefore $\abscentrality_{_Q}(C_{i})$ as well, in $\mathcal{O}(n^2)$.
\end{IEEEproof}

\bigskip
\subsection{Proposition~\ref{prop:fast-updates-same-step}}
\label{appe:fast-updates-same-step}

\begin{proposition-non}
Let $C$ be a set of absorbing nodes, $\mathbf{P}$ the corresponding
transition matrix, and $\mathbf{F} = (\mathbf{I}-\mathbf{P})^{-1}$.
Let $C' = C - \{v\}  \cup \{u\}$, for $u, v \in C$. 
Given  $\mathbf{F}$, the centrality score
$\abscentrality_Q(C')$ can be computed in time $\mathcal{O}(n^2)$.
\end{proposition-non}
\begin{IEEEproof}
The proof is similar to the proof of Proposition~\ref{prop:fast-updates-next-step}.
Without loss of generality, let the two sets of absorbing nodes be
\begin{eqnarray*}
C & = & \{1,2,\ldots, i-1, i\}, \,\text{ and} \\
C' & = & \{1,2,\ldots, i-1, i+1\}.
\end{eqnarray*}
Let $\mathbf{P'}$ be the transition matrix with absorbing nodes $C'$.
The \absorbingcentrality\ for the two sets
of absorbing nodes $C$ and $C'$ is expressed as a function
of the following two matrices
\begin{equation*}
\mathbf{F} = \mathbf{A}^{-1}, \text{ with } \mathbf{A} = \mathbf{I} - \mathbf{P}, \text{ and}
\end{equation*}
\begin{equation*}
\mathbf{F'} = \mathbf{A'}^{-1}, \text{ with } \mathbf{A'} = (\mathbf{I} - \mathbf{P'}).
\end{equation*}
Notice that
\begin{eqnarray*}
& \mathbf{A'} - \mathbf{A} = 
(\mathbf{I} - \mathbf{P'}) - (\mathbf{I} - \mathbf{P}) =
\mathbf{P} - \mathbf{P'} & \\
&
= \left[
\begin{array}{c}
{\mathbf{0}_{(i-1)\times n}} \\
-p_{i, 1}\ \ldots\ -p_{i, n} \\
p_{i+1, 0} \ldots\ p_{i+1, n} \\
{\mathbf{0}_{(n-i-1)\times n}}
\end{array}
\right]
= \mathbf{a}_2 \mathbf{b}_2^{T} - \mathbf{a}_1 \mathbf{b}_1^{T} &
\end{eqnarray*}
where $p_{i, j}$ denotes the transition probability from node $i$ to node $j$ in
a transition matrix $\mathbf{P}_0$ where neither node $i$ or $i+1$ is
absorbing, and the column-vectors $\mathbf{a}_1$, $\mathbf{b}_1$, $\mathbf{a}_2$, $\mathbf{b}_2$ 
are defined as
\begin{eqnarray*}
\mathbf{a}_1 & = & [\overbrace{0 \ldots 0}^{i-1} \ 1 \ 0 \ \overbrace{0 \ldots 0}^{n-i-1}] \\
\mathbf{b}_1 & = & [p_{i, 1}\ \ldots\ p_{i, n}] \\
\mathbf{a}_2 & = & [\overbrace{0 \ldots 0}^{i-1} \ 0 \ 1 \ \overbrace{0 \ldots 0}^{n-i-1}] \\
\mathbf{b}_2 & = & [p_{i+1, 1} \ldots\ p_{i+1, n}].
\end{eqnarray*}
By an argument similar with the one we made in the proof of 
Proposition~\ref{prop:fast-updates-next-step},
we can compute $\mathbf{F'}$ in the following two steps from $\mathbf{F}$, each costing $\mathcal{O}(n^2)$
operations for the provided parenthesization
\begin{eqnarray*}
\mathbf{Z} & = & \mathbf{F}  - (\mathbf{Z} \mathbf{a}_2)(\mathbf{b}_2^{T} \mathbf{Z}) 
/ (1 + \mathbf{b}_2^{T} (\mathbf{Z} \mathbf{a}_2)), \\
\mathbf{F'} & = &  \mathbf{Z} + (\mathbf{F} \mathbf{a}_1)(\mathbf{b}_1^{T} \mathbf{F}) 
/ (1 + \mathbf{b}_1^{T}(\mathbf{F}\mathbf{a}_1)).
\end{eqnarray*}
We have thus shown that, given $\mathbf{F}$, we can compute $\mathbf{F'}$,
and therefore $\abscentrality_{_Q}(C')$ as well, in time $\mathcal{O}(n^2)$.
\end{IEEEproof}


\end{document}

%% file: macros.tex
\usepackage{cite}
\usepackage{url}
\usepackage[pdfauthor={Charalampos Mavroforakis}, pdftitle={Absorbing random-walk centrality - theory and algorithms}, hidelinks, bookmarks=false]{hyperref}
\usepackage{epsfig}
\usepackage{amssymb}
\usepackage{amsmath}
\usepackage{amsfonts}
\usepackage{float}
\usepackage{xcolor, colortbl}
\usepackage{graphicx}
\DeclareGraphicsExtensions{.pdf,.png}
\usepackage{dsfont}
\let\labelindent\relax
\usepackage{enumitem}
\usepackage[scanall]{psfrag}
\usepackage{ntheorem}
\usepackage{multirow}
\usepackage{times}
\usepackage{algorithm}
\usepackage[noend]{algorithmic}
\usepackage{booktabs}

\usepackage{caption}
\usepackage{subcaption}

\newtheorem{theorem}{Theorem}

\newtheorem{definition}{Definition}
\newtheorem{lemma}{Lemma}

\newtheorem{problem}{Problem}

\newtheorem{proposition}{Proposition}
\newtheorem*{proposition-non}{Proposition}

\newcommand{\para}[1]{\noindent{\bf{#1}}}
\newcommand{\sparanbf}[1]{\smallskip\noindent{{#1}}}
\newcommand{\spara}[1]{\smallskip\noindent{\bf{#1}}}

\newcommand{\NP}{{\ensuremath{\mathbf{NP}}}}

\newcommand{\absorbingcentrality}{{{absorbing centrality}}}

\newcommand{\distcentrality}{{\ensuremath{\mathrm{dc}}}}
\newcommand{\abscentrality}{{\ensuremath{\mathrm{ac}}}}
\newcommand{\acg}{{\ensuremath{\mathrm{acg}}}}

\newcommand{\startprob}{{\ensuremath{\mathbf{s}}}}
\newcommand{\m}{{\ensuremath{m}}}
\newcommand{\M}{{\ensuremath{M}}}
\newcommand{\nsph}{\ensuremath{s}}
\newcommand{\stepstoZ}{{\ensuremath{\mathbf{\Lambda}}}}

\newcommand{\true}{{{\tt true}}}

\newcommand{\krwc}{{{$k$-{\sc arw}-{\sc Central}\-{\sc ity}}}}

\newcommand{\VC}{{\sc VertexCover}}

\newcommand{\greedy}{{{\sf\small Greedy}}}
\newcommand{\approxac}{{{\sf\small  Approximate\-AC}}}

\newcommand{\degree}{{{\sf\small  Degree}}}
\newcommand{\closest}{{{\sf\small  Distance}}}
\newcommand{\spectralqueries}{{{\sf\small  SpectralQ}}}
\newcommand{\spectralcandidates}{{{\sf\small  SpectralC}}}
\newcommand{\spectraldistance}{{{\sf\small  SpectralD}}}
\newcommand{\personalpagerank}{{{\sf\small  PPR}}}

\newcommand{\karate}{{{\tt karate}}}
\newcommand{\dolphins}{{{\tt dolphins}}}
\newcommand{\lesmis}{{{\tt lesmis}}}
\newcommand{\adjnoun}{{{\tt adjnoun}}}
\newcommand{\football}{{{\tt football}}}
\newcommand{\livejournal}{{{\tt livejournal}}}
\newcommand{\dblpcoauthors}{{{\tt kddCoauthors}}}
\newcommand{\oregon}{{{\tt oregon-1}}}
\newcommand{\cagrqc}{{{\tt ca-GrQc}}}
\newcommand{\roadnet}{{{\tt roadnet}}}
\newcommand{\cahepth}{{{\tt ca-HepTh}}}

\newcommand{\squishlist}{\begin{list}{$\bullet$}
  { \setlength{\itemsep}{0pt}
     \setlength{\parsep}{3pt}
     \setlength{\topsep}{3pt}
     \setlength{\partopsep}{0pt}
     \setlength{\leftmargin}{1.5em}
     \setlength{\labelwidth}{1em}
     \setlength{\labelsep}{0.5em} } }
\newcommand{\squishend}{
\end{list}  }